\documentclass{article}

\usepackage{arxiv}

\usepackage[utf8]{inputenc} % allow utf-8 input
\usepackage[T1]{fontenc}    % use 8-bit T1 fonts
\usepackage{lmodern}        % https://github.com/rstudio/rticles/issues/343
\usepackage{hyperref}       % hyperlinks
\usepackage{url}            % simple URL typesetting
\usepackage{booktabs}       % professional-quality tables
\usepackage{amsfonts}       % blackboard math symbols
\usepackage{nicefrac}       % compact symbols for 1/2, etc.
\usepackage{microtype}      % microtypography
\usepackage{graphicx}

\title{Citation Analysis of Computer Systems Papers}

\author{ Eitan Frachtenber, Reed College (eitan@reed.edu) }

\providecommand{\tightlist}{%
  \setlength{\itemsep}{0pt}\setlength{\parskip}{0pt}}

\usepackage{longtable,booktabs,array}
\usepackage{calc} % for calculating minipage widths
\usepackage{etoolbox}
\makeatletter
\patchcmd\longtable{\par}{\if@noskipsec\mbox{}\fi\par}{}{}
\makeatother
\IfFileExists{footnotehyper.sty}{\usepackage{footnotehyper}}{\usepackage{footnote}}
\makesavenoteenv{longtable}

\begin{document}
\maketitle

\begin{abstract}
Citation analysis is used extensively in the bibliometrics literature to assess the impact of individual works, researchers, institutions, and even entire fields of study.
In this paper, we analyze citations in one large and influential field within computer science, namely computer systems. Using citation data from a cross-sectional sample of 2,088 papers in 50 systems conferences from 2017, we examine four research questions: overall distribution of systems citations; their evolution over time; the differences between databases (Google Scholar and Scopus) for systems papers, and; the characteristics of self-citations in the field.

We find that only 1.5\% of papers remain uncited after five years, while 12.8\% accrued at least 100 citations, both statistics comparing favorably to many other scientific fields.
The most cited subfields and conference areas within systems were security, databases, and computer architecture.
Most papers achieved their first citation within a year from publication, and the median citation count continued to grow at an almost linear rate over five years, with only a few papers peaking before that.
We also find that early citations could be linked to papers with a freely available preprint, or may be primarily composed of self-citations.
The ratio of self-citations to total citations starts relatively high for most papers, but appears to stabilize by 12--18 months, at which point highly cited papers revert to predominately external citations.
Past self-citation count (taken from each paper's reference list) appears to bear little if any relationship with the future self-citation count of each paper.
The choice of citation database also makes little difference in relative citation comparisons, despite marked differences in absolute counts.
\end{abstract}


\hypertarget{introduction}{%
\section{Introduction}\label{introduction}}

Citation analysis plays a central role in bibliometric evaluation of journals, conferences, institutes, and individual researchers (Moed 2006).
The advent of web-based citation databases has led to faster growth in their importance and use (Meho 2007).
Nonetheless, citation analysis remains challenging when comparing citations across years (Varga 2019), types of scholarly communication (Martins et al. 2010), or fields of study (Adam 2002; Patience et al. 2017).

In this study, we explore a multifaceted citation analysis that circumvents these challenges by focusing on papers from a single year, from conferences only, and from a single large field within computer science (CS), namely, computer systems (or just ``systems'' for short).

The bibliometrics literature is rich with studies analyzing citations in various disciplines and fields, including CS as a whole (Devarakonda et al. 2020; Hirst and Talent 1977; Mattauch et al. 2020).
Even within CS, several fields, subfields, and specific venues have received bibliometric analyses (Broch 2001; Frachtenberg 2022b; Iqbal, Qadir, Hassan, et al. 2019; Iqbal, Qadir, Tyson, et al. 2019; Lister and Box 2008; Rahm and Thor 2005; Wang et al. 2016).
To the best of our knowledge, this study is the first systematic analysis of a wide cross-section of computer systems research.

Systems is a large research field with numerous applications, used by some of the largest technology companies in the world.
For the purpose of this study, we define systems as the study and engineering of concrete computing systems, which includes research topics such as operating systems, computer architectures, data storage and management, compilers, parallel and distributed computing, and computer networks.
The goal of this study is to characterize, for the first time, the citation behavior of this large and important CS field, while controlling for year, venue type, and research area.
Specifically, we examine: total citations after five years and how they compare across subfields and against other disciplines and fields; the dynamics of citation counts over time; the effect of the source of the citation database; and the characteristics of self-citations in computer systems.

\hypertarget{study-design-and-main-findings}{%
\subsubsection*{Study design and main findings}\label{study-design-and-main-findings}}
\addcontentsline{toc}{subsubsection}{Study design and main findings}

This study uses an observational, cross-sectional approach, analyzing 2,088 papers from a large subset of leading systems conferences.
The study population came from a hand-curated collection of 50 peer-reviewed systems conferences from a single publication year (2017).
Among other characteristics, the dataset includes paper citation counts at regular intervals for at least five years from the publication date, by which time many citation statistics stabilize (Larivière, Gingras, and Archambault 2009).
Using this dataset, this study addresses the following high-level research questions:

\textbf{RQ1: What is the distribution of paper citations in systems conferences after five years?}
We find the typical skewed distributions with most papers accruing a few citations, a handful of papers racking up thousands of citations, and a dearth of uncited papers. According to one comparison, citation counts of top-cited papers put the field of systems among the top ten scientific fields. However, citations are distributed unevenly across and within conferences and subfields, with security, databases, and computer architecture conferences ranking the highest median citations.

\textbf{RQ2: How have citations evolved over this period?}
In our dataset, the median number of citations per paper grows at a nearly constant rate, with few papers peaking before the five-year mark, and few papers achieving ``runaway'' citation growth.
Most papers are first cited within 9--12 months since publications, and the average time to have to wait to reach at least \(n\) citations grows almost linearly with \(n\).

\textbf{RQ3: How do Google Scholar and Scopus compare for these conferences?}
We find, as have other studies before, that the Google's inclusive policy of counting many document types as potential citation sources does inflate the absolute citation counts compared to Scopus' counts.
However, the two counts are nearly perfectly correlated and provide similar relative comparisons across papers, conferences, and years.

\textbf{RQ4: How many citations are self-citations, and how do they evolve over time?}
The main finding emerging from the dataset is that self-citations are more prevalent among early citations and among papers that are less cited overall after five years.
However, the average ratio of self-citations to total citations appears to stabilize for most papers about 12--18 months after publication.
This dataset shows no clear relationship between a paper's eventual self-citations and the number of self-citations it contains in its own reference list.

\hypertarget{contributions}{%
\subsubsection*{Contributions}\label{contributions}}
\addcontentsline{toc}{subsubsection}{Contributions}

In addition to answering these research questions, this study makes the following contributions:

\begin{itemize}
\item
  A critical view of H-index as a metric for conferences or journals.
\item
  A discussion of the relationship between acceptance rates and citations.
\item
  A characterization of papers that are cited relatively early.
\end{itemize}

As a final contribution, this study provides the dataset of papers and citations over time, tagged with rich metadata from multiple sources (Frachtenberg 2021).
Since comprehensive data on papers and conferences with citations are not always readily available, owing to the significant manual data collection involved, this dataset can serve as the basis of additional studies.

\hypertarget{organization}{%
\subsubsection*{Organization}\label{organization}}
\addcontentsline{toc}{subsubsection}{Organization}

The remainder of this paper is organized as follows.
In the next section (Section \ref{sec:methods}), we describe the data collection and processing methodology in detail.
In the results section (Section \ref{sec:results}), we enumerate our findings, organized by research question.
In Section \ref{sec:discussion}, we combine results on raw citation statistics to explore three higher-level topics.
Related work is surveyed in Section \ref{sec:related}, and in Section \ref{sec:conclusion}, we summarize our results and suggest directions for future research.

\hypertarget{sec:methods}{%
\section{Materials and Methods}\label{sec:methods}}

The most time-consuming aspect of this study was the collection and cleaning of data.
This section describes the data selection and cleaning process for conference, paper, and citation data.

The dataset we analyze comes from a hand-curated collection of 50 peer-reviewed systems conferences from a single publication year (2017) to reduce time-related variance.
Conference papers were preferred over journal articles because in CS, and in particular, in its more applied fields, such as systems, original scientific results are typically first published in peer-reviewed conferences (Patterson, Snyder, and Ullman 1999; Patterson 2004), and then possibly in archival journals, sometimes years later (Vrettas and Sanderson 2015).
These conferences (detailed in Appendix A) were selected to represent a large cross-section of the field with different sizes, competitiveness, and subfields.
Such choices are necessarily subjective, based on the author's experience in the field.
But they are aspirationally both spread enough to represent the field well and focused enough to distinguish it from the rest of CS.
For each conference, we gathered various statistics from its web page, proceedings, or directly from its program chairs, and downloaded all papers in PDF format.

Since we are primarily interested in measuring the posthoc impact of each paper, as approximated by its number of citations, we regularly collected citation data from two databases, Google Scholar (GS) and Scopus.
GS is an extensive database with excellent coverage of CS conferences that contains not only peer-reviewed papers, but also preprints, patents, technical reports, and other sources of unverified quality (Halevi, Moed, and Bar-Ilan 2017).
Consequently, its citation counts tend to be higher than those of databases such as Scopus and Web of Science, but not necessarily inferior when used in paper-to-paper comparisons (Harzing and Alakangas 2016; Martin-Martin et al. 2018).
Since we are mostly comparing relative citation metrics, even if the GS metrics appear inflated compared to other databases, we should still be able to examine the relationship between relative citation counts of papers and conferences.
Nevertheless, for papers covered in the Scopus database, we compare both sources of citations in RQ3 to ensure that both metrics are in relative agreement with each other.

\hypertarget{statistics}{%
\subsection*{Statistics}\label{statistics}}
\addcontentsline{toc}{subsection}{Statistics}

For hypothesis testing, group means were compared pairwise using Welch's two-sample t-test and group medians using the Wilcoxon signed-rank test; differences between distributions of two categorical variables were tested with the \(\chi^{2}\) test; and correlations between two numerical variables were evaluated using Pearson's product-moment correlation coefficient.
All statistical tests are reported with their p-values.
All computations were performed using the R programming language and can be found in the source code accompanying this paper.

\hypertarget{ethics-statement}{%
\subsection*{Ethics statement}\label{ethics-statement}}
\addcontentsline{toc}{subsection}{Ethics statement}

All data for this study were collected from public online sources and therefore did not require the informed consent of the authors.
No funding was received to assist with the preparation of this manuscript.

\hypertarget{code-and-data-availability}{%
\subsection*{Code and data availability}\label{code-and-data-availability}}
\addcontentsline{toc}{subsection}{Code and data availability}

The complete dataset and metadata are available online (Frachtenberg 2021).

\hypertarget{limitations}{%
\subsection*{Limitations}\label{limitations}}
\addcontentsline{toc}{subsection}{Limitations}

To control for the effect of time on citations, we opted not to include additional data from more recent conference years in our dataset. Undoubtedly, more data could strengthen the statistical validity of our observations; but it could also weaken any conclusions based on the inherent delays in the citation process and in variation over time.
Our methodology is also constrained by the manual collection of data, such as conference statistics; paper downloading and text conversion; cleanup and verification; etc.
The effort involved in compiling all necessary data limits the scalability of our approach to additional conferences or years.

Our focus on citations as a primary metric of interest has also received significant criticism in the bibliometrics literature because of its own limitations.
First, there are limitations stemming from the sources of citation data, such as inflated metrics in GS and partial coverage in Scopus, as previously mentioned.
Second, even accurate citation counts present a variety of limitations, such as references expressing negative sentiments (Parthasarathy and Tomar 2014) or researchers and venues gaming their own citation counts (Biagioli 2016).
Nevertheless, citations are arguably the most popular method to evaluate research impact, either as raw counts or as inputs to compound metrics, such as the H-index.
As such, this initial analysis of citation behavior in a previously unexplored and important field could further our understanding of both the particular field and bibliometrics as a whole.

\hypertarget{sec:results}{%
\section{Results}\label{sec:results}}

\hypertarget{rq1-citation-distribution}{%
\subsection{RQ1: Citation distribution}\label{rq1-citation-distribution}}

We begin by describing the distribution of citations in the field.
In addition to observing summary statistics, such as means, medians, and outliers, focusing our attention on distributions can increase transparency when comparing citation metrics because of their typical skewness (Larivière et al. 2016).
To better understand the distribution of 5-year citations, we start top-down by looking at the overall distribution and characteristics of all cited papers and then zoom in on distributions by conference and by paper topics.
We then fill in the missing piece by describing the distribution of uncited papers.

\hypertarget{overall-distribution}{%
\subsubsection{Overall distribution}\label{overall-distribution}}

\begin{figure}
\includegraphics[width=0.75\textwidth]{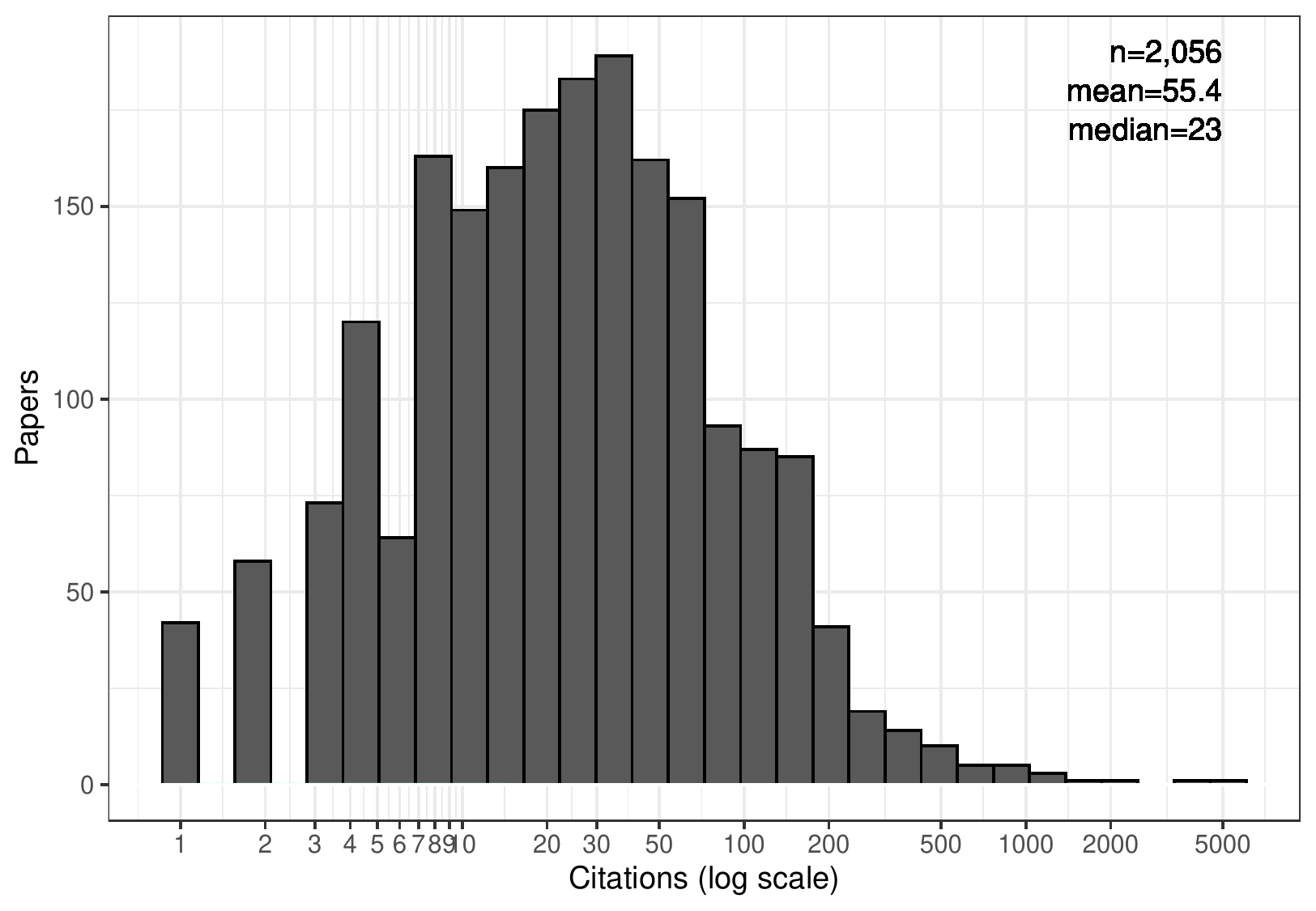} \caption{citation distribution after five years of all cited papers. Also shown are the number of samples (papers), mean, and median citations per paper.}\label{fig:citation-histogram}
\end{figure}

Figure \ref{fig:citation-histogram} shows the overall distribution of cited papers exactly five years from their publication date.
It is indeed skewed and long-tailed (note the logarithmic scale) and resembles a log-normal distribution (Rahm and Thor 2005; Redner 1998; Wang and Barabási 2021; Wu, Luesukprasert, and Lee 2009).
Consequently, the mean number of citations, 55, is much higher than the median,
23, and even higher than the \(75^{th}\) percentile, 53.
The mean is pushed this high by a handful of outlier papers: there are just 7 papers with over 1,000 citations each, one as high as 5229!
Similarly, we find as many as 267 papers (or
12.8\% of the dataset) with at least 100 citations.
This large group of papers would probably no longer be considered an outlier, and 100 citations are likely a reliable signal of notable impact, again suggesting that the field as a whole may be quite influential.

If we follow Patience's method (Patience et al. 2017) to attenuate the outliers and compute the mean number of five-year citations among the top-cited 31--500 papers, we get an average of 120.
That same study also compared this statistic across 236 science categories.
So, using their data for comparison would rank systems as one of the top ten categories in citation count (but keep in mind that the original comparison used less-inflated citations counts from the Web of Science database.)

\hypertarget{citations-by-conference}{%
\subsubsection{Citations by conference}\label{citations-by-conference}}

\begin{figure}
\centering
\includegraphics{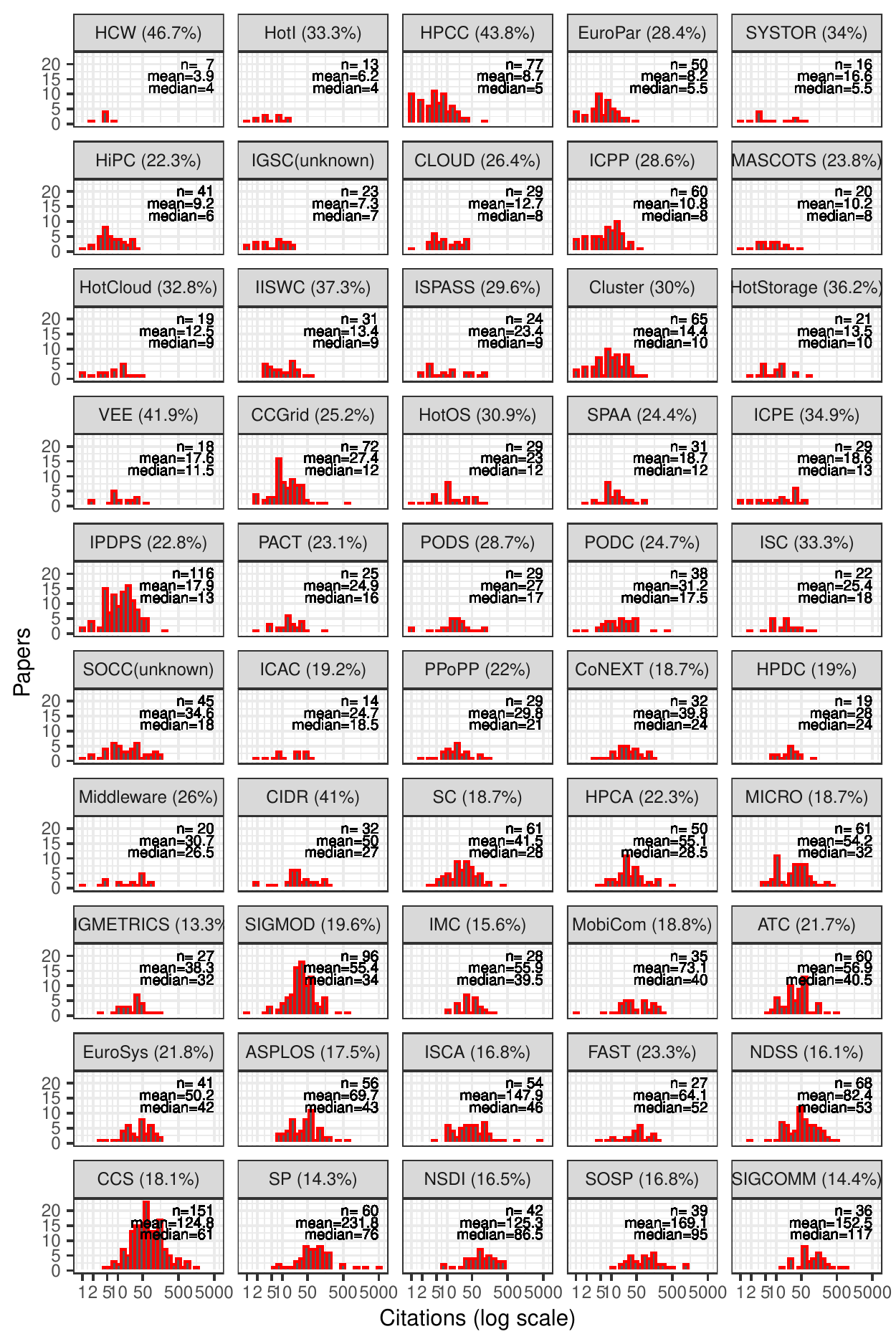}
\caption{\label{fig:citation-histogram-by-conf}Citation distribution of all cited papers by conference. Also shown are the number of samples (papers), mean, and median citations per paper. Conferences ordered by median and showing acceptance rate in parenthesis next to their names.}
\end{figure}

Although overall citation counts for computer systems appear high compared to other fields, keep in mind that not all conferences within the field are equally well cited.
Figure \ref{fig:citation-histogram-by-conf} shows the same kind of histogram, but this time broken down by conference and sorted by median citation per conference.
We can observe the following relationships from the data:

\begin{itemize}
\tightlist
\item
  Citations vary widely between conferences, ranging from an average of about 4 for HCW papers to about 152 for SIGCOMM papers.
\item
  Even within conferences, some outlier papers push the mean far from the median. For example, SP and ISCA exhibit long tails, with a few papers having thousands of citations, whereas NDSS, an equally well-cited conference, exhibits nearly identical mean and median.
\item
  The likelihood of a conference containing an outlier paper with at least 500 citations increases with conference size. Papers with over 500 citations were published in conferences averaging 85.7 accepted papers, vs.~an average of 58.6 for papers with fewer than 500 citations (\(t=2.47\), \(p=0.02\)). Furthermore, a conference's size is weakly but positively correlated with its average paper citations (\(r=0.32\), \(p=0.03\)), suggesting a random variable in the citation count of a paper.
\item
  Median citations, on the other hand, do not appear to be significantly correlated with a conference's number of accepted papers (\(r=0.24\), \(p=0.09\)). However, conferences with higher median citations are generally more competitive (i.e., lower acceptance rates). These two factors exhibit a strong negative correlation (\(r=-0.64\), \(p<10^{-5}\)). Likewise, mean citations per conference is also negatively correlated with the conference's acceptance rate (\(r=-0.59\), \(p<10^{-5}\)).
\item
  Different conferences have different skews: the distribution can be wide or narrow; some have long right tails and some have none; and the modes appear at different locations, not always near the median. In other words, despite all conferences belonging to the same overall systems field, citation distributions still vary significantly with other factors including size, acceptance rate, and specific subfield.
\item
  Related, some conferences sharing subfields of systems appear to be also clustered together in citation metrics. For example, NDSS, CCS, and SP---all focused on computer security---are similarly ranked in terms of median citations (although the means vary much more). Likewise, NSDI/SIGCOMM (focusing on networking), HPCA/MICRO and ASPLOS/ISCA (focusing on computer architecture) also pair with similar median citations.
\end{itemize}

This last observation naturally leads to the question of the relationship between a research subfield and its typical citations.
To answer this question, we next look at research topics at the individual paper level rather than at the conference's broad scope.

\hypertarget{citations-by-subfield}{%
\subsubsection{citations by Subfield}\label{citations-by-subfield}}

The definition of subfields in computer systems, like the definition of the field itself, is necessarily fluid and subjective.
Based on research experience in the field, we made a best-effort attempt to categorize all papers by reading every single abstract and tagging each paper with one or more subfields.
The list of selected tags is presented in Table \ref{tab:topic-tags}.
Naturally, experts would differ in their opinions on this list and possibly on tag assignments to papers, but the current assignment provides a starting point for comparison across systems subfields.

\begin{table}

\caption{\label{tab:topic-tags}List of selects systems subfields}
\centering
\begin{tabular}[t]{ll}
\toprule
Tag & Subfield description\\
\midrule
Architecture & Computer architecture\\
Benchmark & Workloads, bechnmarking, metrics, and methodology\\
Cloud & Cloud computing and related infrastructure\\
Compilers & Compilers, language and debugging sopport, runtime systems\\
Concurrency & Parallel and distributed computing\\
Data & Big data applications and infrastructure\\
DB & Databases, key-value stores, and database management systems\\
Energy & Power and energy efficiency, sustainable computing\\
GPGPU & Accelerator technologies and heterogeneous computing\\
HPC & High performance computing and supercomputing applications\\
Network & Networking algorithms, wireless networks, switching and routing\\
OS & Operating systems, scheduling, resource management\\
Security & Security, privacy, encryption, resilience\\
Storage & Persistent and ehpemeral storage: disk, flash, RAM, etc.\\
VM & Containers and virtual machines and, networks\\
\bottomrule
\end{tabular}
\end{table}

\begin{figure}
\centering
\includegraphics{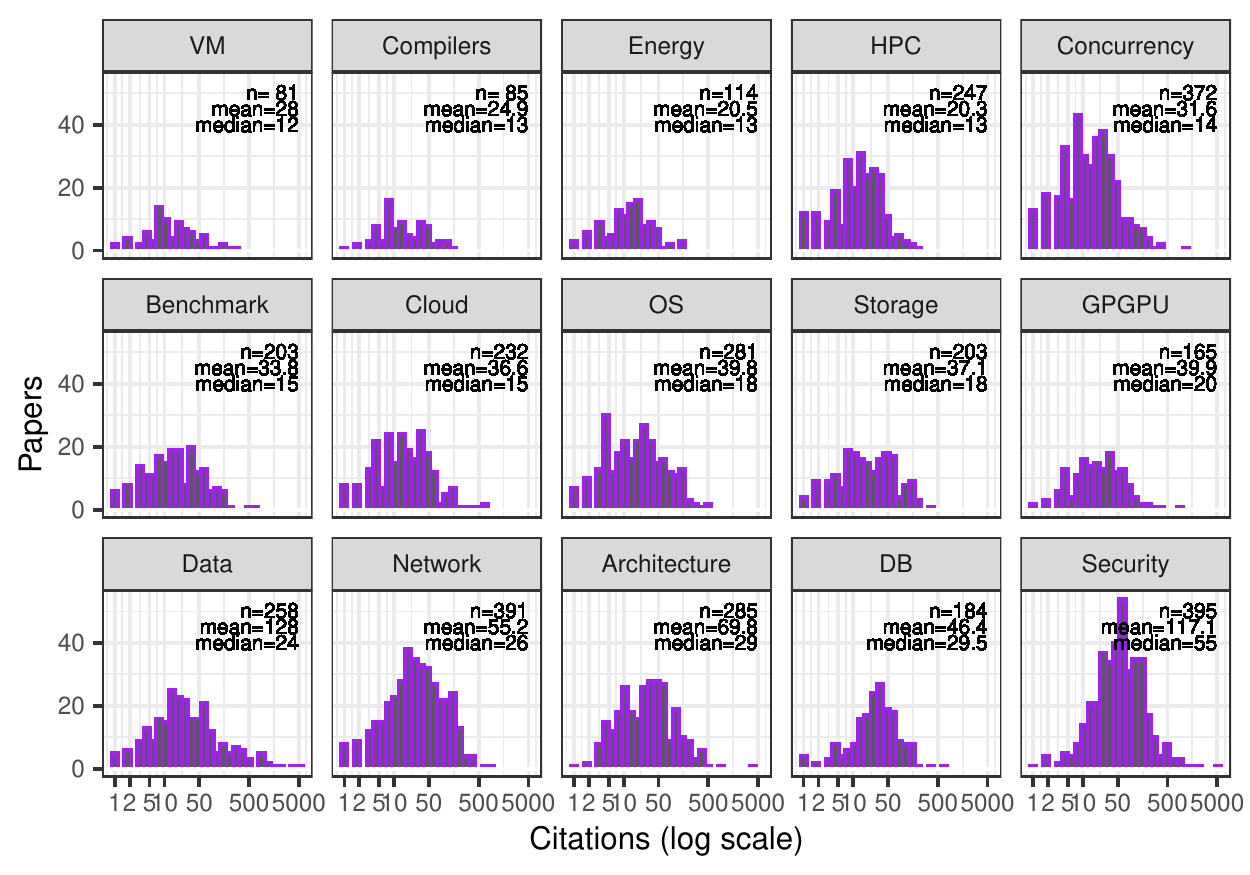}
\caption{\label{fig:citation-by-topic}Citation distribution for all cited papers by paper's topic. Papers with multiple topics appear in multiple histograms. Also shown are the number of samples (papers), mean, and median citations per paper. Topics ordered by median citations per paper.}
\end{figure}

Papers in our dataset are generally focused on one or two of these topics, averaging
1.7 topics per paper.
There are a total of 2,026 (97\% of all papers)
cited papers with at least one topic assigned.
The remainder either had no citations or did not fit well with any of the tags.
Figure \ref{fig:citation-by-topic} shows the distribution of citations by topic in these tagged papers (with multitopic papers repeated).

Again, there appears to be a positive relationship between quantity---the number of papers published on a topic---and ``quality''---how well cited these papers generally are (\(r=0.57\), \(p=0.03\)).
This could again indicate a random element leading to a higher probability of outlier papers in large subfields, or it could indicate that these topics were both sufficiently popular at the time to attract multiple submissions and popular over time to attract eventual citations.

Nevertheless, this correlation should also be considered with a grain of salt because of the subjective process of conference selection for this study.
For example, if more conferences and workshops on virtualization had been chosen, the number of papers on the topic would have obviously increased, whereas the relatively large conferences on security skew topic counts from the other end.
At any rate, in 2017, the broad research topics that resulted in the highest median citations appear to be security, databases and data management, computer architecture, and networks.

\hypertarget{uncited-papers}{%
\subsubsection{Uncited papers}\label{uncited-papers}}

Owing to the logarithmic citation scale, the preceding histograms omitted papers with zero citations.
Some early studies claimed that, generally, most scientific papers are not cited at all (Hamilton 1991; Jacques and Sebire 2010; Meho 2007).
More recent research found that the rate of uncited papers keeps decreasing, and estimates it to be less than 10\% (Wu, Luesukprasert, and Lee 2009).
For example, one study computed the percentage of uncited papers in physics (11\%), chemistry (8\%), and biomedical sciences (4\%) (Larivière, Gingras, and Archambault 2009).
Another large-scale estimate for the entire CS discipline found that 44.8\% of CS papers remained uncited after five years (Chakraborty et al. 2015).
In our dataset of systems papers, only 32 papers remain uncited after five years
(1.5\%).
Of these, the conference with the most uncited papers was HPCC, followed by IPDPS and EuroPar (Table \ref{tab:uncited-conf}).
Both in absolute terms and in percentage terms, the number of uncited papers remains very low for most conferences.

\begin{table}

\caption{\label{tab:uncited-conf}Uncited papers by conference, as paper count and percentage of all accepted papers.}
\centering
\begin{tabular}[t]{lrr}
\toprule
Conference & Count & Percent\\
\midrule
HPCC & 6 & 7.79\%\\
IPDPS & 4 & 3.45\%\\
EuroPar & 3 & 6\%\\
Cluster & 2 & 3.08\%\\
HiPC & 2 & 4.88\%\\
ICPE & 2 & 6.9\%\\
ISPASS & 2 & 8.33\%\\
CCGrid & 1 & 1.39\%\\
CLOUD & 1 & 3.45\%\\
HCW & 1 & 14.29\%\\
HotStorage & 1 & 4.76\%\\
HPCA & 1 & 2\%\\
HPDC & 1 & 5.26\%\\
ICPP & 1 & 1.67\%\\
IGSC & 1 & 4.35\%\\
IISWC & 1 & 3.23\%\\
PODC & 1 & 2.63\%\\
SYSTOR & 1 & 6.25\%\\
\bottomrule
\end{tabular}
\end{table}

\begin{table}

\caption{\label{tab:uncited-topics}Uncited papers by topic tags}
\centering
\begin{tabular}[t]{ll}
\toprule
Topic & Count\\
\midrule
Benchmark & 8\\
HPC & 7\\
OS & 7\\
Storage & 7\\
Concurrency & 6\\
GPGPU & 6\\
Data & 5\\
Network & 4\\
Energy & 2\\
Architecture & 1\\
Cloud & 1\\
Compilers & 1\\
\bottomrule
\end{tabular}
\end{table}

The topic tags with the most uncited papers were Benchmark, Storage, and HPC, followed by GPGPU and Concurrency (Table \ref{tab:uncited-topics}).
It is not surprising that the two distributions appear to be related.
Many papers in the top six uncited conferences were tagged with some of the top six uncited topics.

Having examined the citation distribution at a fixed point in time, we can now examine how it evolved to this point from the date of publication.

\hypertarget{rq2-citation-dynamics}{%
\subsection{RQ2: Citation dynamics}\label{rq2-citation-dynamics}}

Observing the total citations of papers at a fixed time point offers only a static view of a metric that is inherently a moving target.
Citations tend to follow different dynamics as different papers, disciplines, and fields exhibit very different aging curves (Pichappan and Ponnudurai 1999; Wang 2013).
After five years, all the papers in our dataset likely had a chance to be discovered by fellow researchers, as evidenced by the fact that nearly all are cited by outside researchers.
We can therefore concentrate next on citation dynamics.

For our second research question, we examine three aspects of the time dimension: general distribution over time, time to first citation, and citation velocity.
Since different conferences publish in different months, causing nearly a year's gap between the first and the last, we normalize time for all three aspects by counting the number of months passed from the date it was published, rather than a fixed start day in 2017.

\hypertarget{citation-distribution}{%
\subsubsection{Citation distribution}\label{citation-distribution}}

\begin{figure}
\includegraphics[width=0.75\textwidth]{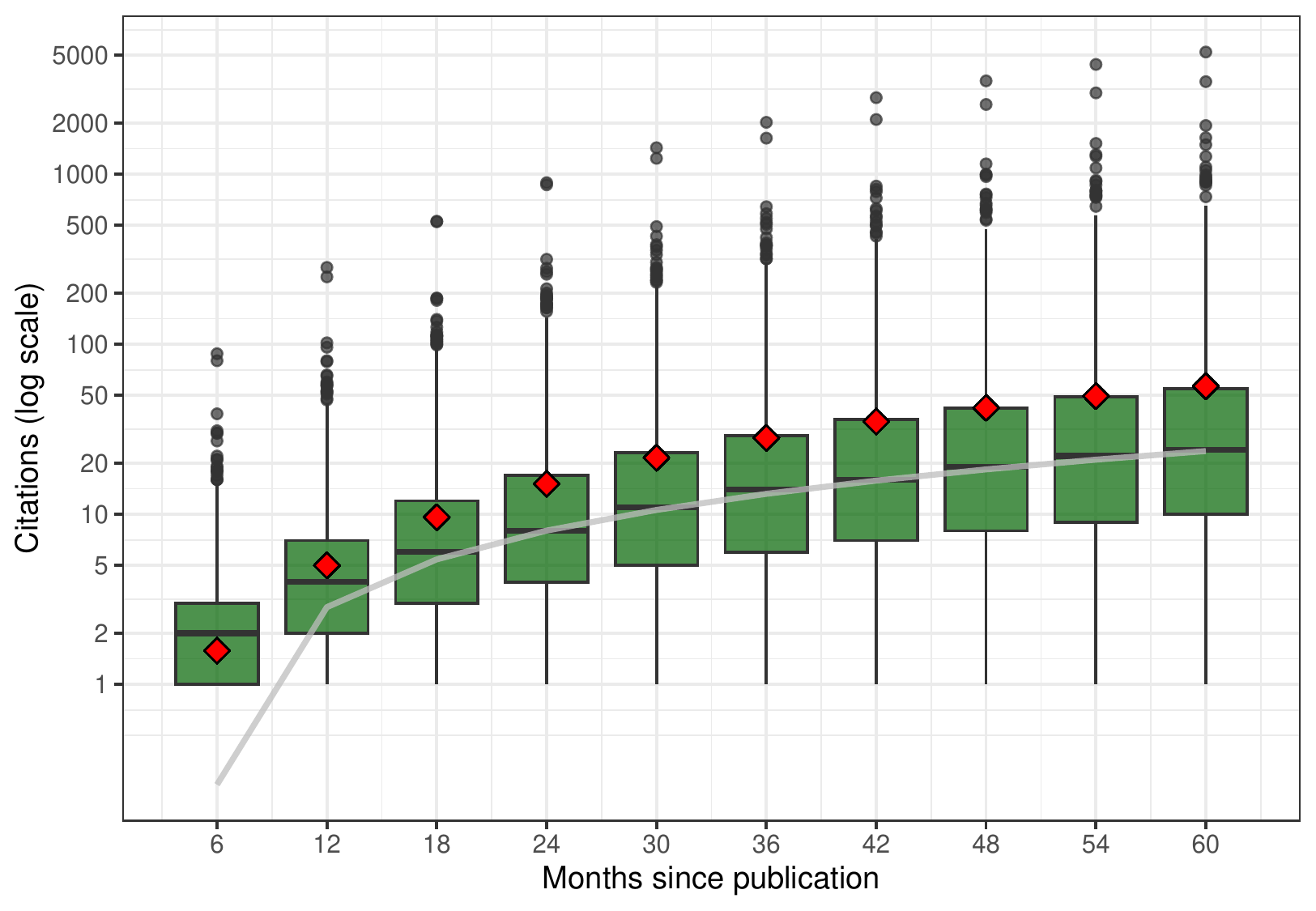} \caption{Citation distribution for all cited papers over time. Center bars show median value; lower and upper hinges correspond to the 25th and 75th percentile; upper points show high-citation outliers; and diamonds show means for all citations, including uncited papers. Also shown in gray is a linear regression line for the median values.}\label{fig:citation-over-time}
\end{figure}

Looking at the citation distribution at six-month intervals (Figure \ref{fig:citation-over-time}), three observations can be made.
First, we can see that the spread---difference across papers---grows over time (note the logarithmic scale), as some papers accelerate at a faster rate than others, creating a larger range of citation values.

overall citations, as expressed by the 25th, 50th, and 75th quantiles grow at a rapid clip over time without showing signs of slowing down.
In fact, if we fit the median with a simple linear regression model, shown as a gray line,
we match the median citation values after the first year nearly perfectly
(intercept: -2.33, slope: 0.43 citations per month).
Linear models for the 25th and 75th percentiles would have different slopes, because of the increasing spreads, but still fit quite well.

In contrast, mean citation soutpace medians' growth because of the disproportionate pull of outlier papers.
Except for the first six months (when the difference between the mean and the median is less than half a citation), the presence of highly cited papers pushes the mean increasingly higher than the median.
In the first six months, the presence of many
(1,191)
uncited papers is pulling the mean to
\(\approx{}\) 1.6
citations.
However, as papers gain enough time to be discovered, cited, peer-reviewed, and then published, the mean quickly catches up.
By 12 months, the mean number of uncited papers drops to
503, and by 18 months to
227.

This observation naturally leads us to the next aspect of time: how long does it take papers to be cited?

\hypertarget{time-to-first-citation}{%
\subsubsection{Time to first citation}\label{time-to-first-citation}}

\begin{figure}
\includegraphics[width=0.75\textwidth]{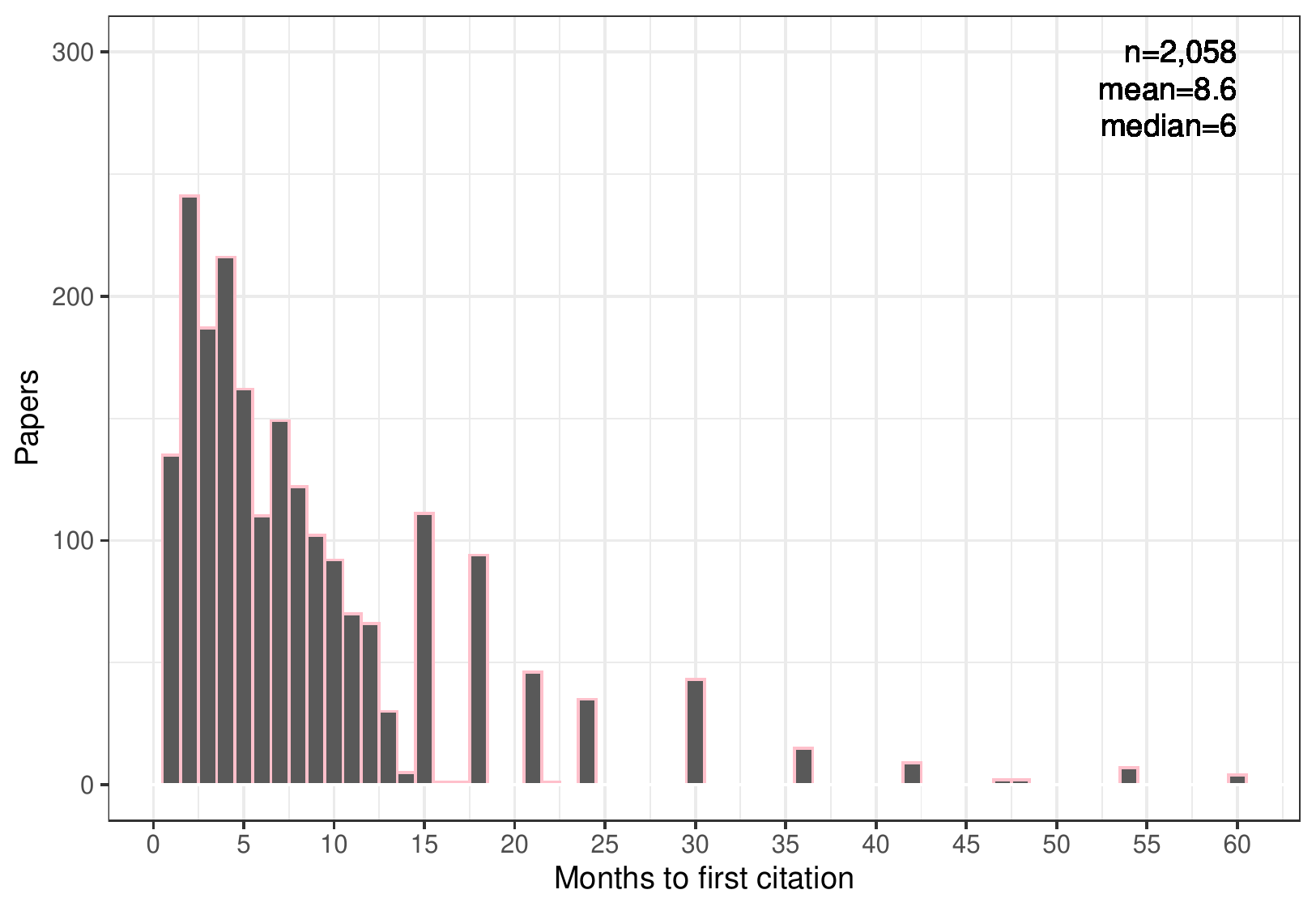} \caption{Distribution of time to first citation for all cited papers. Also shown are the number of samples (papers), mean, and median times in months.}\label{fig:first-citation-histogram}
\end{figure}

Our dataset includes citation data at monthly intervals during the first year from publication, then quarterly for the second year, and then every six months for the remaining three years.
This resolution allows us to estimate with near-monthly accuracy the first time that GS detected a citation for each paper.
Figure \ref{fig:first-citation-histogram} shows the distribution of the approximate time in months it took GS to first detect any citations for each paper (compared to Figure \ref{fig:citation-histogram}, two papers briefly showed citations before reverting back to zero).

Assuming that the first citation is external, i.e., not a self-citation, we can think of this moment of first being cited as a paper's ``discovery event.''
In our dataset, it averages about 9 months.
(As we see later in RQ4, this assumption is invalid for many papers, which may postpone the discovery time by a few months.)
Discovery time can be longer than subsequent citations because it requires the paper to be published and discovered by another researcher, who must then wait till their citing document is peer-reviewed and published, a process that can take several months.

\begin{figure}
\includegraphics[width=0.75\textwidth]{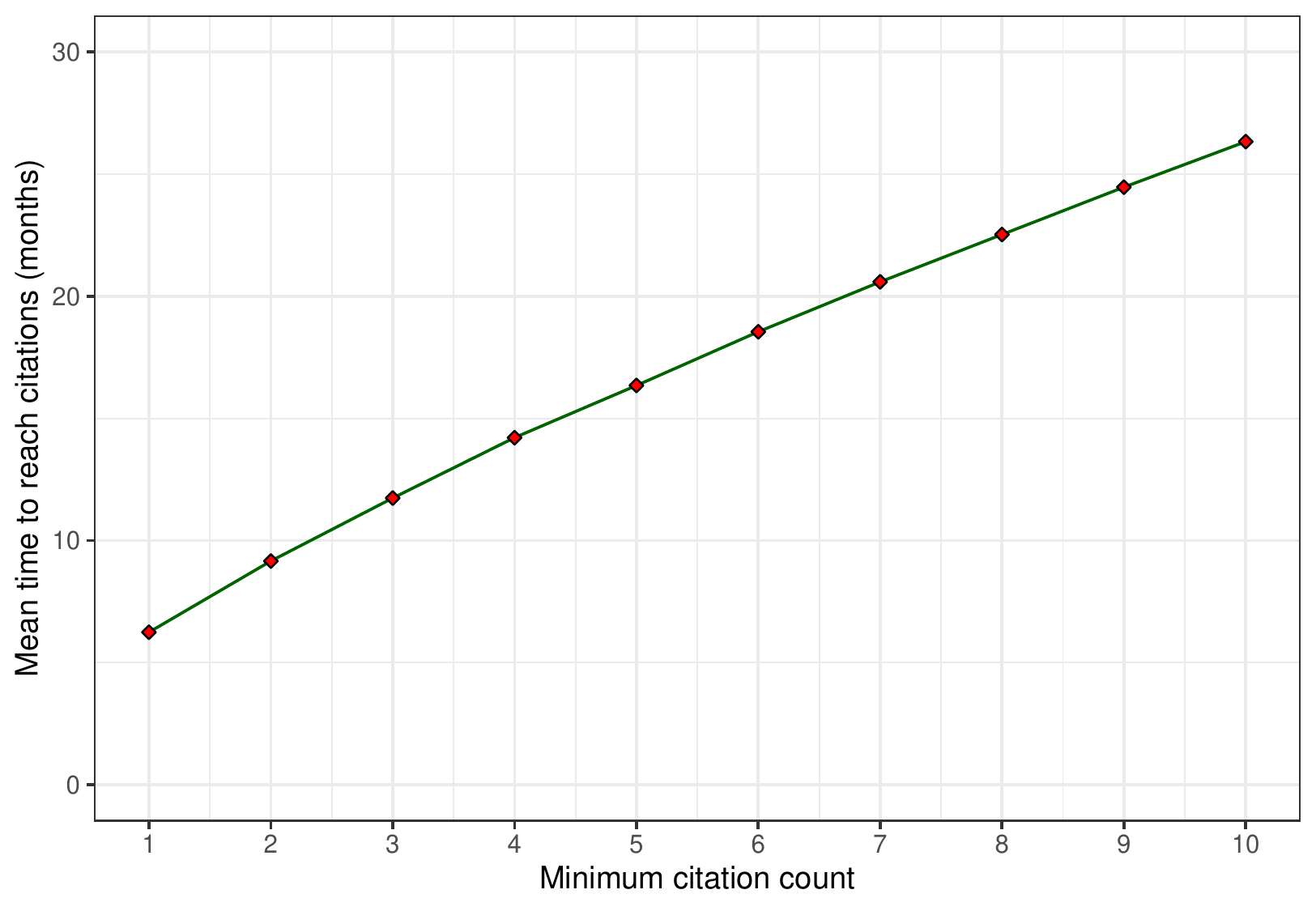} \caption{Average time to reach n citations for increasing values of n. Only papers with at least 10 citations after five years are included.}\label{fig:time-to-n-citations}
\end{figure}

To illustrate this point, Figure \ref{fig:time-to-n-citations} shows the mean time it takes for papers to reach \texttt{n} citations.
Note that the average growth from the first citation to the next and then the next takes about the same time across the range (about 2--3 months for each additional citation, slowly decreasing).
However, going from zero citations to the first citation takes on average about twice as long.
This observation leads us to the next and final question of time, namely, how fast and how long do citations grow?

\hypertarget{citation-velocity}{%
\subsubsection{Citation velocity}\label{citation-velocity}}

\begin{figure}
\centering
\includegraphics{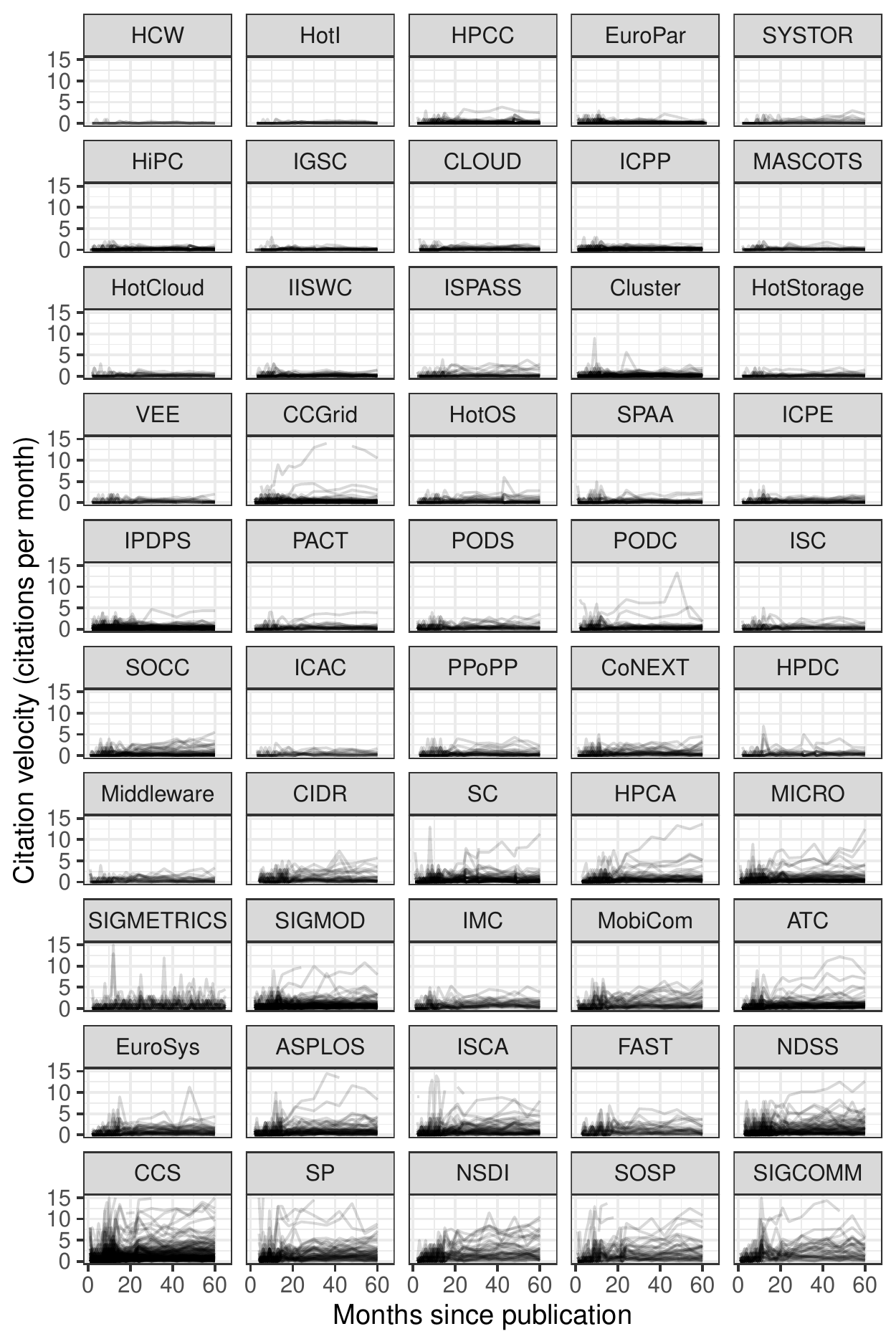}
\caption{\label{fig:citation-velocity}Citation velocity of every paper. Y-axis is clipped to show the bulk of the data with higher resolution, but some papers exceed 100 citations per month. Conferences ordered again by increasing median citations after five years.}
\end{figure}

As we have just observed, the mean time across papers to receive the first citation
(6.24 months)
is longer than the time to add the second citation
(2.91 months).
It continues to decrease slowly such that it only takes
1.86 months on average to add the \(10^{th}\) citation.
Intuitively, this makes sense as papers are discovered by an increasingly larger network of researchers and pick up more and more citations during their growth phase.
However, this phase cannot grow indefinitely, as the size of the potential network is bounded by the systems research community size, and the impact of systems papers is often limited to a few years until newer systems replace them.
To visualize these growth patterns, we can plot the citation velocity of every paper (Figure \ref{fig:citation-velocity}).

The first observation we can make from the figure is that the citation growth for most papers is nearly flat.
It hovers close to zero for conferences with low median citations and remains mostly below five citations per month even for better-cited conferences.
The second observation is that a few papers do achieve runaway growth, sometimes exceeding the plot limit of 15 citations per month---even reaching 147 in one instance---but their overall number is small.
Only 26 papers exceed ten citations per month at the end of the five years.

This brings us to the last observation and provides an answer to the question we first posed.
There were actually slightly more papers exceeding ten citations per month at the four-year mark (30 papers), meaning that a few of those peaked before reaching five years.
Looking at the graphs (ignoring some sharp peaks and dips caused by noise from GS), we can observe the occasional paper that reached maximum velocity and then started to slow down.
Again, the overall number of these appears small, suggesting that most papers in systems take longer than five years to peak.

\hypertarget{rq3-comparison-of-gs-and-scopus-citations}{%
\subsection{RQ3: Comparison of GS and Scopus citations}\label{rq3-comparison-of-gs-and-scopus-citations}}

The citation criteria employed by GS tend to be more inclusive than those of other databases, sometimes resulting in inflated citation counts (Harzing and Alakangas 2016; Martin-Martin et al. 2018).
Moreover, different fields tend to be covered to different degrees by different databases, and not all databases cover conferences equally well (we did get 100\% coverage for our systems conferences in GS).
Because of this criticism of GS and as a way to add a measure of control to our findings, we can examine citation data from another database, Springer's Scopus.

We collected two types of statistics from Scopus: total citations exactly five years from each paper's publication and total citations at the end of each year (with and without self-citations).
We first focus on the former statistic, which is directly comparable to the one we collected from GS, and
explore the latter in the next research question.

\begin{figure}
\includegraphics[width=0.75\textwidth]{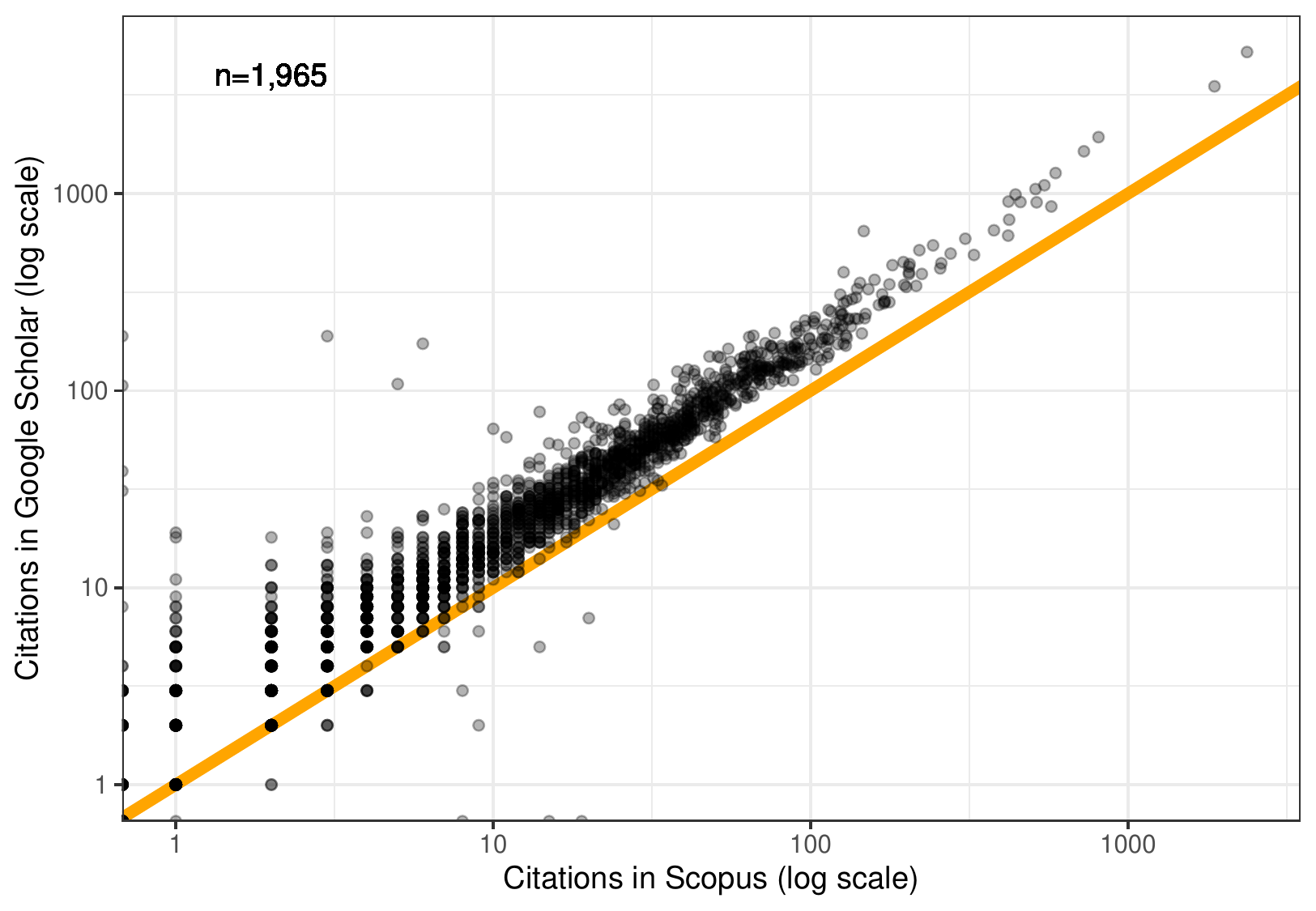} \caption{Comparison of GS and Scopus citations after five years. Orange line denotes a 1:1 mapping.}\label{fig:gs-vs-scopus}
\end{figure}

Note that Scopus coverage for these systems conferences is incomplete, missing a total of 123 papers relative to GS.
For the remaining 1965 papers, Figure \ref{fig:gs-vs-scopus} shows a scatter plot of every paper's citation count in Scopus (x-axis) and GS (y-axis).
The vast majority of points lie above the 1:1 intercept line, confirming that GS citation counts tend to be higher than those of Scopus.
But the difference in magnitudes appears to be remarkably constant, with near-perfect correlation
(\(r=0.99\), \(p<10^{-9}\), \(R^{2}=\) 0.98).
The implication here is that all the relative observations we have drawn so far on GS citations should generalize to Scopus citations as well, up to a constant factor.
Other studies have found that Google Scholar citations are strongly correlated with those from Web of Science as well (Kousha and Thelwall 2007).

\hypertarget{rq4-effect-of-self-citations}{%
\subsection{RQ4: Effect of self-citations}\label{rq4-effect-of-self-citations}}

Self-citations are fairly common in the sciences and have been estimated to comprise 10--40\% of all scientific production, depending on the field (Aksnes 2003; Snyder and Bonzi 1998; Wolfgang, Bart, and Balázs 2004).
On the one hand, self-citations represent a natural evolution of a research team's work, building upon their previous results, especially in systems projects that often involve incremental efforts of implementation, measurement, and analysis (Wolfgang, Bart, and Balázs 2004).
On the other hand, self-citations can be problematic as a bibliometric measure of a work's impact because they obscure the external reception of the work and are prone to manipulation (Waltman 2016).

For our final research question, we would like to quantify the degree to which self-citations affect the overall citation metrics in computer systems and how well self-citations can be predicted from the papers' references themselves.
To this end, we examine how self-citations evolve over time, how they relate to self-citations in the original papers, and how common they are overall.

\hypertarget{ratio-of-self-citations-over-time}{%
\subsubsection{Ratio of self-citations over time}\label{ratio-of-self-citations-over-time}}

\begin{figure}
\includegraphics[width=0.75\textwidth]{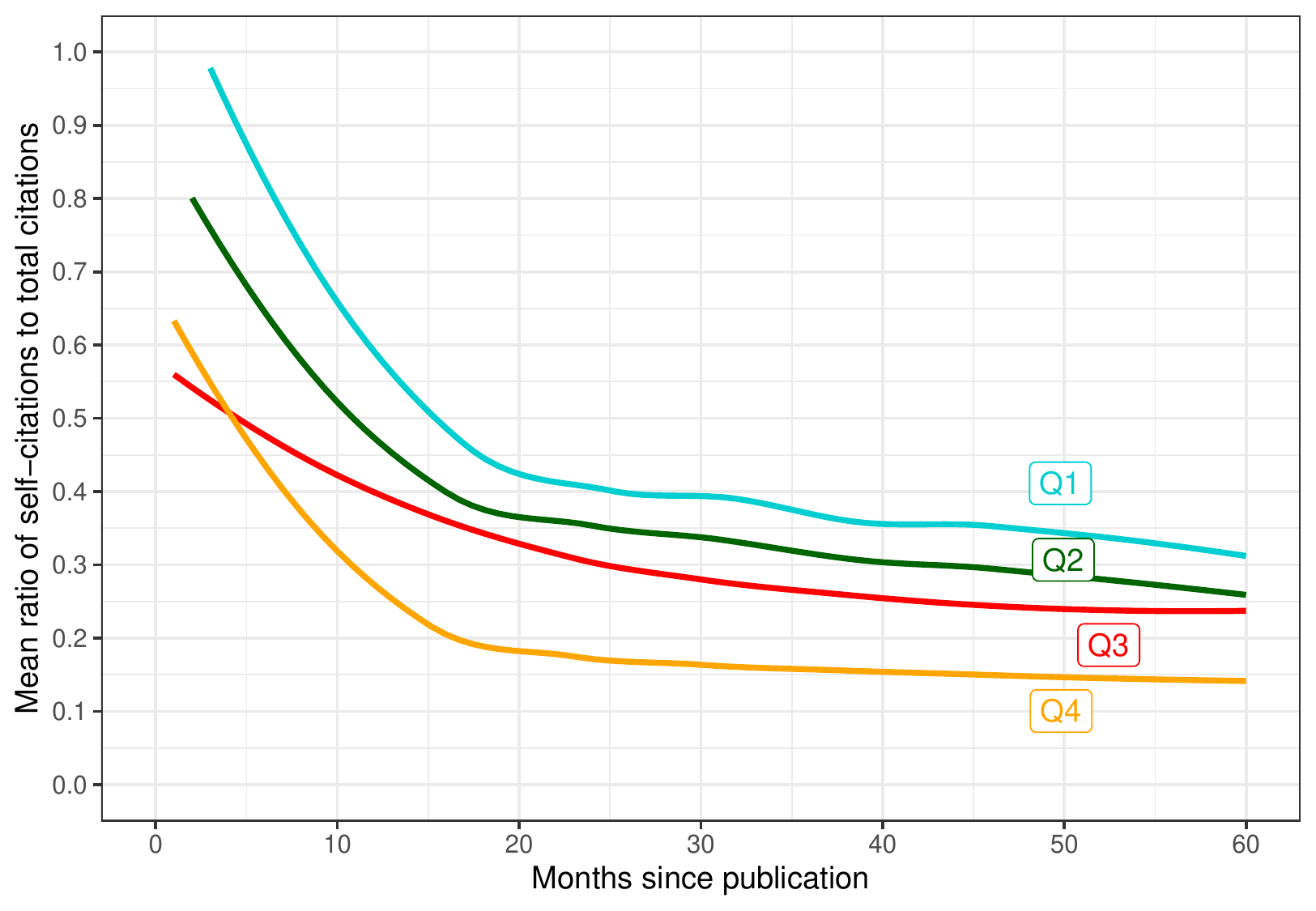} \caption{Mean ratio of self-citations out of total citations as a function of months since publication. Trend lines are smoothed using LOESS. Data are divided into four quartiles based on how many total citations the paper had after 5 years (Q4 means most cited papers).}\label{fig:self-citation-ratio}
\end{figure}

We start by computing the ratio of self-citations for every paper at every measurement time point (the end of each calendar year between 2017 and 2022, which represents different ages in months for each paper depending on the month they were published in).
We then average these ratios across all papers with the same measurement age, and divide those into four quartiles based on the total citations they had at the time.
The results of this computation are shown in Figure \ref{fig:self-citation-ratio}.
A few notable observations emerged from the data:

\begin{itemize}
\tightlist
\item
  In the first few months after publication, most of the detected citations appear to be self-citations for all groups. A distinct transition to majority external citations starts about a year after publication.
\item
  Highly cited papers (e.g., Q3 and Q4) are also cited earlier.
\item
  The more a paper is cited overall, the lower the ratio of self-citations. Conversely, for Q1 and Q2, nearly all early citations were self-citations.
\item
  That said, even in the fourth quartile, self-citations still comprise some 10\% of all five-year citations on average.
\end{itemize}

Overall, it appears that self-citations are more characteristic of early citations, either because a paper has not had enough time to be well-known by almost anyone other than its authors, or as an attempt by the authors to increase its visibility (Chakraborty et al. 2015).
But as time passes and papers are evaluated by the community, highly cited papers accrue citations primarily from external researchers.

\hypertarget{backward-and-forward-self-citations}{%
\subsubsection{Backward and forward self-citations}\label{backward-and-forward-self-citations}}

These self-citation distributions inspire another curious question: Is there a pattern of specific self-citation behavior that affects both forward and backward citations?
In other words, can we predict the number of self-citations a paper would receive after five years by counting self-citations in its own bibliography?

To answer this question, we estimated the total number of backward self-citations, as follows:
First, we converted the reference list from each PDF file to a text file for every single paper and manually cleaned up conversion artifacts such as two-column papers breaking down the reference list.
Then, we searched for the last names of all authors in every single reference.
Any reference that matched one or more names was counted as a self-citation.
Clearly, this method is imperfect because some last names may match different authors, and some last names may even represent a word in a paper's title.
Nevertheless, manual inspection of several dozen papers has revealed very few inaccuracies.

\begin{figure}
\includegraphics[width=0.75\textwidth]{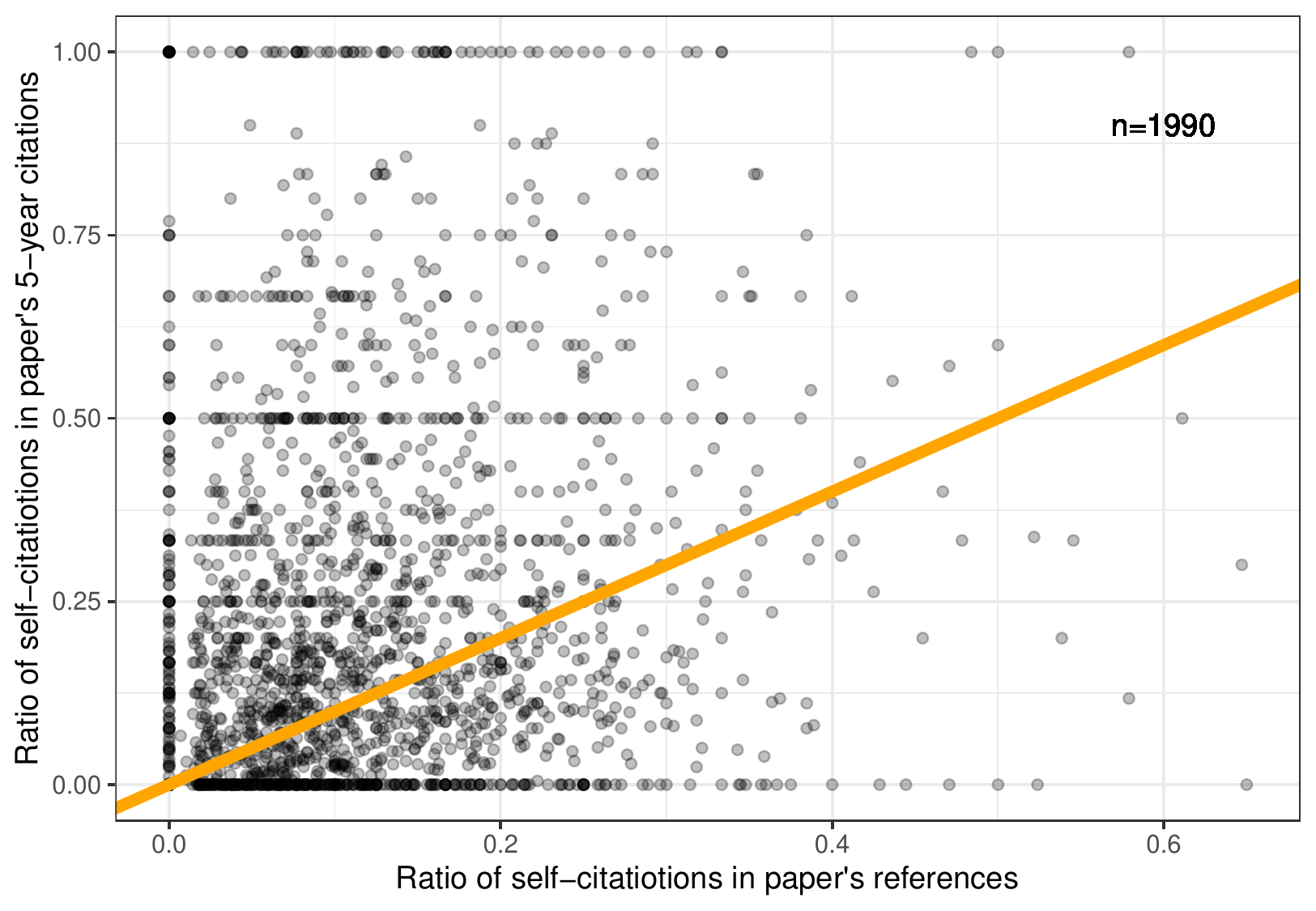} \caption{Relationship between five-year self-citations and self-citation count in each paper's own references. Orange line denotes a  1:1 mapping.}\label{fig:ref-correlations}
\end{figure}

The self-reference counts were converted to ratios by dividing by the total number of references in each paper and then compared to eventual (5-year) self-citations.
Figure \ref{fig:ref-correlations} shows the relationship between these two measures.
We can make out several horizontal bands for low-cited papers with fixed ratios of eventual citations (0\%, 50\%, 100\%, etc.), and one clear vertical band for papers with zero backward self-references.
But overall, it shows little discernible patterns.

Likewise, the Pearson correlation between backward and forward self-citation ratios is nominal
(\(r=0.16\), \(p<10^{-9}\)).
Removing some of the outliers (the top and bottom horizontal bands, as well as the leftmost band) does little to increase the correlation
(\(r=0.19\), \(p<10^{-9}\)).
Even if we look only at the temporal subset of data when papers first exceed five citations (which typically comprise of a higher ratio of self-citations),
the observed correlation is not particularly strong
(\(r=0.24\), \(p<10^{-9}\)).
It appears that we reached a mostly negative result for our question.

\hypertarget{overall-self-citation-ratios}{%
\subsubsection{Overall self-citation ratios}\label{overall-self-citation-ratios}}

One anecdotal observation we can make from Figure \ref{fig:ref-correlations} is that most papers appear above the orange 1:1 line.
In other words, a majority of papers
(1,136 or
57.1\%
to be exact) had a higher self-citation ratio after five years than their backward self-citation ratio in their reference list.

Moreover, the mean self-citation ratio after five years,
24.1\%,
is on the one hand higher than the 15\% rate found for physical sciences (and 9\% over all sciences) in an earlier study on self-citation patterns across disciplines (Snyder and Bonzi 1998); but on the other hand, is in close agreement with the 24\% rate found for CS papers in Norway (Aksnes 2003).
The Norway study also found a high ratio of self-citing papers overall, agreeing with our data where 78.4\%
of papers had at least one self-citation after five years.

In terms of outliers, hidden in these ratios and averages are several papers whose total citations were predominantly self-citations.
In their final citation count, 252 papers had over 50\% self-citations, of which 11 papers were in Q4 with dozens of citations each.
Four of the six most self-cited papers in relative terms were in architecture conferences, suggesting that research in this area often builds primarily on past research from the same group.
That said, the overall correlation between a paper's total citations and self-citations after five years is low
(\(r=0.19\), \(p<10^{-9}\)),
suggesting that for this dataset as a whole, self-citations are not typically a dominant component of overall citations.

\hypertarget{sec:discussion}{%
\section{Discussion}\label{sec:discussion}}

In this section, we integrate results from the previous section to explore three topics beyond raw citation counts: the use of H-index as a measure of conference quality, the relationship between a conference's acceptance rate and its eventual citations, and the characteristics of papers with early citations.

\hypertarget{conference-h-index}{%
\subsection{Conference H-index}\label{conference-h-index}}

One derivative aspect of citations that we can calculate is a conference's H-index.
H-index was developed by Hirsch to measure the impact of individual researchers (Hirsch 2005).
It is defined as the maximum integer \(n\) such that there exist \(n\) papers published in a given time window that received at least \(n\) citations.
GS also reports the H-index and five-year H-index of conferences and journals and ranks them accordingly (Patience et al. 2017).
With our dataset of 2017 papers, we can compute a similar measure based on papers from that single year, if nothing else, to illustrate the weakness of this metric when applied to conferences.

Our partial H5-index metric ranges from
4 for HCW to
69 for CCS.
It correlates strongly with the size of the conference
(\(r=0.75\), \(p<10^{-9}\)),
suggesting that a conference's mere number of accepted papers, largely a policy decision, has a significant impact on its H-index metric.
Moreover, recall the observation that the larger a conference, the more likely it is to have a ``runaway'' successful paper.
This suggests that among all the factors that impact a conference's total citations (and consequently its H-index), there is also a non-negligible element of chance.
In other words, a conference's steering committee could conceivably increase its H-index (and average citations) by simply increasing the size of the conference.
Obviously, size is not the most important factor, and an overly permissive acceptance policy is likely to admit papers that would lower the overall citation average.
As demonstrated in our dataset, the partial H5-index metric is clearly negatively correlated with acceptance rate
(\(r=-0.62\), \(p<10^{-5}\)).
Nevertheless, this finding weakens the case for the use of these metrics as reliable indicators of conference quality.

\hypertarget{conference-acceptance-rate}{%
\subsection{Conference acceptance rate}\label{conference-acceptance-rate}}

The strong correlation between a conference's acceptance rate and its median citation counts could suggest a causal relationship, that is, that competitive peer review selects for high-impact papers.
However, the peer-review process is notoriously unreliable when it comes to selecting highly cited papers, which may not always even be the reviewers' goal (Coupé 2013; Lee 2019; Wainer, Eckmann, and Rocha 2015).
Moreover, even if a causal relationship does exist, its direction is unclear.
That is to say, researchers who predict that their paper will have a relatively low citation impact may self-select to submit it to a less competitive conference to increase its chances of acceptance.
Since we regrettably cannot design a randomized controlled trial where some papers are randomly accepted in a given conference, and since we cannot accurately predict the citation impact of rejected papers, we do not have the necessary tools to evaluate the strength and direction of any such causal link.
That said, even if the peer-review process has little predictive or selective power for high-impact papers, we could surmise that successfully publishing in competitive conferences increases the likelihood of eventual citations because intuitively, prestigious conferences increase the post-publication visibility and credibility of their papers.

\hypertarget{early-cited-papers}{%
\subsection{Early cited papers}\label{early-cited-papers}}

Although we expect several months to pass before a paper is first discovered and cited by external scientists,
some ``early'' papers were cited within six months of publication.
Our dataset provides evidence for two possible explanations.

First, many of these citations could be self-citations, which do not require discovery.
As we have seen for RQ4, more early citations are indeed self-citations.
For example, the
255
``early'' papers that were cited within six months of publication (Scopus data) averaged
55.61\%
self-citations.
In contradistinction, in the
463
papers first cited in the following six-month period (months 7--12), this proportion was significantly lower at
41.98\%
(\(t=3.79\), \(p<10^{-3}\)).

Second, the availability of preprints or other freely accessible versions of the paper \emph{before} publication could accelerate the discovery process.
We measured the time (in months) it took GS to discover a freely available e-print version of each paper.
For the ``early'' papers, this time averaged
3.76
months, compared to
5.55
months for the slow papers
(\(t=-3.19\), \(p<0.01\)).
For a frame of reference, this average was close to the overall average time to e-print across all papers
(5.47 months).
From this perspective, therefore,
the papers cited within the second half-year are indistinguishable from all later-to-cite papers, as opposed to the distinct ``early'' papers cited in the first half-year.

\hypertarget{sec:related}{%
\section{Related work}\label{sec:related}}

Citation analysis is an active area of research in bibliometrics, with many studies looking at it both quantitatively---examining citation distributions, as this study does---or qualitatively, examining the strengths, weaknesses, and characteristics of citations as a metric of scholarly impact.
This study does not aim to debate the merits of citation-based metrics, so we will not review here the qualitative aspects.
The interested reader is referred to the recent book ``The Science of Science'' for an overview of this debate (Wang and Barabási 2021).

On the quantitative side, we find several studies that examine some of the metrics we discussed in this paper, as well as others.
Tsay and Shu analyzed citations in one journal and found that the most cited documents were journal articles, followed by books and book chapters, electronic resources, and conference proceedings (Tsay and Shu 2011).
As mentioned in the introduction, in computer science, conference papers take precedence to journal articles (Goodrum et al. 2001; Vrettas and Sanderson 2015).

Citation analyses collect data from different databases, primarily Google Scholar, Scopus, and Web of Science.
Other studies have evaluated these databases for their validity and reliability in citation analysis.
Several of those concluded that Google Scholar both has higher coverage and more liberal policies for defining what constitutes a citation (Halevi, Moed, and Bar-Ilan 2017), resulting in higher citation counts overall, as we have also found (Kousha and Thelwall 2007).
This difference can make it difficult to argue about absolute numbers of citations since they vary significantly by database.
However, as most of these studies also found---and ours agrees---the citation counts from the three databases appear to be strongly correlated (Harzing and Alakangas 2016; Martin-Martin et al. 2018).
The implication is that comparing citations across papers, conferences, and years should produce similar conclusions, regardless of which reputable database is used for citation data.

Numerous studies examined specific aspects of citation analysis, such as the type of citing document or type of cited document.
For example, Harter measured the citation counts and impact factor of early electronic journals to compare them with traditional journals (Harter 1996).
A related study looked at how often electronic resources are cited in electronic journals (Herring 2002).
Various other studies have found a potential link between the open sharing of research artifacts and increased citations (Frachtenberg 2022a; Heumüller et al. 2020).

There also exist many citation analyses for various disciplines and venues, too numerous to enumerate here.
Of particular relevance, however, we can find similar studies for other fields of computer science (Broch 2001; Lister and Box 2008; Wang et al. 2016).
Even specific systems subfields and systems conferences were analyzed, such as performance evaluation (Frachtenberg 2022b), networking (Iqbal, Qadir, Hassan, et al. 2019; Iqbal, Qadir, Tyson, et al. 2019), cloud computing (Khan, Arjmandi, and Yuvaraj 2022), and databases (Rahm and Thor 2005).
To the best of our knowledge, this study is the first to analyze citation characteristics across the entire field of computer systems research.

\hypertarget{sec:conclusion}{%
\section{Conclusion and future work}\label{sec:conclusion}}

This work examined, for the first time, the citation characteristics of an important field of computer science, namely computer systems.
It found that compared to other fields and disciplines, systems papers are very well cited: the top-cited papers rank the field among the highest scientific disciplines in citation counts, and only a few papers remain uncited after five years.

The overall ratio of five-year self-citations (24.1\%) appears to be higher than in other scientific fields but agrees with the 24\% rate found for all computer science papers in Norway.
Interestingly, most papers exhibit some self-citations, especially within the first few months since publication (when the free availability of e-print versions of a paper also increases its early citation count).
Over time, the self-citation ratio remains above 30\% for papers with relatively few citations, while for the most cited 25\% of papers, this ratio drops below 10\%.

Competitive conferences with a low acceptance rate tend to higher citations per paper on average.
On the other hand, conferences that accept many papers overall (independent of the acceptance ratio) also exhibit more ``runaway success'' papers that lift their average citation counts, suggesting the possibility of a random factor.
It also suggests that the H5-index metric for conferences is particularly prone to manipulation, as increasing the number of accepted papers is correlated with higher H5-index values on average.
Whether the citation count comes from Google Scholar or Scopus did not matter much, as both were strongly correlated (Pearson's r=0.99), leading to similar findings for all relative comparisons across papers and conferences.

This work can be extended in several directions.
Two obvious extensions along the time axis are to continue collecting citation metrics for the same conferences and to look at additional conference years in order to verify the generalization of these results.
A more intricate proposal is to move from the ``how many''-type questions on citations in this study to the ``why''-type questions.
The goal of such an investigation would be to identify which among dozens of factors are most closely associated with increased citation counts in the field of computer systems and hypothesize potential explanations.
We plan to follow up with a study that analyzes the myriad factors that could be associated with citation counts: conference-related, paper-related, and author-related.

\hypertarget{appendix-a.-detailed-conference-list}{%
\section*{Appendix A. Detailed conference list}\label{appendix-a.-detailed-conference-list}}
\addcontentsline{toc}{section}{Appendix A. Detailed conference list}

Each conference is described by its initialism, full name, commencement date, size (number of published papers), acceptance rate (if known), and the homepage containing the program.

\begin{enumerate}
\item ASPLOS: ACM International Conference on Architectural Support for Programming Languages and Operating Systems, 2017-04-08. 56 papers, acceptance rate: 17.5\%. Homepage: \url{http://novel.ict.ac.cn/ASPLOS2017/}
\item ATC: USENIX Annual Technical Conference, 2017-07-12. 60 papers, acceptance rate: 21.7\%. Homepage: \url{https://www.usenix.org/conference/atc17}
\item CCGrid: IEEE/ACM CCGrid, 2017-05-14. 72 papers, acceptance rate: 25.2\%. Homepage: \url{https://www.arcos.inf.uc3m.es/wp/ccgrid2017/}
\item CCS: ACM Conference on Computer and Communications Security, 2017-10-31. 151 papers, acceptance rate: 18.1\%. Homepage: \url{https://www.sigsac.org/ccs/CCS2017/}
\item CIDR: The biennial Conference on Innovative Data Systems Research, 2017-01-08. 32 papers, acceptance rate: 41\%. Homepage: \url{http://cidrdb.org/cidr2017/}
\item CLOUD: IEEE International Conference on Cloud Computing, 2017-06-25. 29 papers, acceptance rate: 26.4\%. Homepage: \url{http://www.thecloudcomputing.org/2017/}
\item Cluster: IEEE Cluster Conference, 2017-09-05. 65 papers, acceptance rate: 30\%. Homepage: \url{https://cluster17.github.io/}
\item CoNEXT: ACM International Conference on Emerging Networking Experiments and Technologies, 2017-12-13. 32 papers, acceptance rate: 18.7\%. Homepage: \url{http://conferences2.sigcomm.org/co-next/2017/#!/home}
\item EuroPar: International European Conference on Parallel and Distributed Computing, 2017-08-30. 50 papers, acceptance rate: 28.4\%. Homepage: \url{http://europar2017.usc.es/}
\item EuroSys: The European Conference on Computer Systems, 2017-04-23. 41 papers, acceptance rate: 21.8\%. Homepage: \url{https://eurosys2017.github.io/}
\item FAST: USENIX Conference on File and Storage Technologies, 2017-02-27. 27 papers, acceptance rate: 23.3\%. Homepage: \url{https://www.usenix.org/conference/fast17/}
\item HCW: IEEE International Heterogeneity in Computing Workshop, 2017-05-29. 7 papers, acceptance rate: 46.7\%. Homepage: \url{http://hcw.eecs.wsu.edu/}
\item HiPC: IEEE International Conference on High Performance Computing, Data, and Analytics, 2017-12-18. 41 papers, acceptance rate: 22.3\%. Homepage: \url{http://hipc.org/}
\item HotCloud: USENIX Workshop in Hot Topics in Cloud Computing, 2017-07-10. 19 papers, acceptance rate: 32.8\%. Homepage: \url{https://www.usenix.org/conference/hotcloud17}
\item HotI: IEEE Annual Symposium on High-Performance Interconnects, 2017-08-28. 13 papers, acceptance rate: 33.3\%. Homepage: \url{http://www.hoti.org/hoti25/archives/}
\item HotOS: ACM Workshop on Hot Topics in Operating Systems, 2017-05-07. 29 papers, acceptance rate: 30.9\%. Homepage: \url{https://www.sigops.org/hotos/hotos17/}
\item HotStorage: USENIX Workshop on Hot Topics in Storage and File Systems, 2017-07-10. 21 papers, acceptance rate: 36.2\%. Homepage: \url{https://www.usenix.org/conference/hotstorage17}
\item HPCA: The IEEE Symposium on High Performance Computer Architecture, 2017-02-04. 50 papers, acceptance rate: 22.3\%. Homepage: \url{http://hpca2017.org}
\item HPCC: IEEE International Conference on High Performance Computing and Communications, 2017-12-18. 77 papers, acceptance rate: 43.8\%. Homepage: \url{http://hpcl.seas.gwu.edu/hpcc2017/}
\item HPDC: ACM International Symposium on High Performance Parallel and Distributed Computing, 2017-06-28. 19 papers, acceptance rate: 19\%. Homepage: \url{http://www.hpdc.org/2017/}
\item ICAC: IEEE International Conference on Autonomic Computing, 2017-07-18. 14 papers, acceptance rate: 19.2\%. Homepage: \url{http://icac2017.ece.ohio-state.edu/}
\item ICPE: ACM/SPEC International Conference on Performance Engineering, 2017-04-22. 29 papers, acceptance rate: 34.9\%. Homepage: \url{https://icpe2017.spec.org/}
\item ICPP: IEEE International Conference on Parallel Processing, 2017-08-14. 60 papers, acceptance rate: 28.6\%. Homepage: \url{http://www.icpp-conf.org/2017/index.php}
\item IGSC: IEEE International Green and Sustainable Computing Conference, 2017-10-23. 23 papers, acceptance rate: unknown. Homepage: \url{http://igsc.eecs.wsu.edu/}
\item IISWC: IEEE International Symposium on Workload Characterization, 2017-10-02. 31 papers, acceptance rate: 37.3\%. Homepage: \url{http://www.iiswc.org/iiswc2017/index.html}
\item IMC: ACM Internet Measurement Conference, 2017-11-01. 28 papers, acceptance rate: 15.6\%. Homepage: \url{http://conferences.sigcomm.org/imc/2017/}
\item IPDPS: IEEE International Parallel and Distributed Processing Symposium, 2017-05-29. 116 papers, acceptance rate: 22.8\%. Homepage: \url{http://www.ipdps.org/ipdps2017/}
\item ISC: ISC High Performance, 2017-06-18. 22 papers, acceptance rate: 33.3\%. Homepage: \url{http://isc-hpc.com/id-2017.html}
\item ISCA: ACM/IEEEE International Symposium on Computer Architecture, 2017-06-24. 54 papers, acceptance rate: 16.8\%. Homepage: \url{http://isca17.ece.utoronto.ca/doku.php}
\item ISPASS: IEEE International Symposium on Performance Analysis of Systems and Software, 2017-04-24. 24 papers, acceptance rate: 29.6\%. Homepage: \url{http://www.ispass.org/ispass2017/}
\item MASCOTS: IEEE International Symposium on the Modeling, Analysis, and Simulation of Computer and Telecommunication Systems, 2017-09-20. 20 papers, acceptance rate: 23.8\%. Homepage: \url{http://mascots2017.cs.ucalgary.ca/}
\item MICRO: IEEE/ACM International Symposium on Microarchitecture, 2017-10-16. 61 papers, acceptance rate: 18.7\%. Homepage: \url{https://www.microarch.org/micro50/}
\item Middleware: The Annual Middleware Conference, 2017-12-11. 20 papers, acceptance rate: 26\%. Homepage: \url{http://2017.middleware-conference.org/}
\item MobiCom: ACM International Conference on Mobile Computing and Networking, 2017-10-17. 35 papers, acceptance rate: 18.8\%. Homepage: \url{https://sigmobile.org/mobicom/2017/}
\item NDSS: The Network and Distributed System Security Symposium, 2017-02-26. 68 papers, acceptance rate: 16.1\%. Homepage: \url{https://www.ndss-symposium.org/ndss2017/}
\item NSDI: USENIX Symposium on Networked Systems Design and Implementation, 2017-03-27. 42 papers, acceptance rate: 16.5\%. Homepage: \url{https://www.usenix.org/conference/nsdi17/}
\item PACT: IEEE/ACM International Conference on Parallel Architectures and Compilation Techniques, 2017-09-11. 25 papers, acceptance rate: 23.1\%. Homepage: \url{https://parasol.tamu.edu/pact17/}
\item PODC: ACM Symposium on Principles of Distributed Computing, 2017-07-25. 38 papers, acceptance rate: 24.7\%. Homepage: \url{https://www.podc.org/podc2017/}
\item PODS: ACM Symposium on Principles of Database Systems, 2017-05-14. 29 papers, acceptance rate: 28.7\%. Homepage: \url{http://sigmod2017.org/pods-program/}
\item PPoPP: ACM SIGPLAN Symposium on Principles and Practice of Parallel Programming, 2017-02-04. 29 papers, acceptance rate: 22\%. Homepage: \url{http://ppopp17.sigplan.org/}
\item SC: The International Conference for High Performance Computing, Networking, Storage and Analysis, 2017-11-13. 61 papers, acceptance rate: 18.7\%. Homepage: \url{http://sc17.supercomputing.org/}
\item SIGCOMM: ACM SIGCOMM Conference, 2017-08-21. 36 papers, acceptance rate: 14.4\%. Homepage: \url{http://conferences.sigcomm.org/sigcomm/2017/}
\item SIGMETRICS: ACM SIGMETRICS, 2017-06-05. 27 papers, acceptance rate: 13.3\%. Homepage: \url{http://www.sigmetrics.org/sigmetrics2017}
\item SIGMOD: ACM International Conference on Management of Data, 2017-05-14. 96 papers, acceptance rate: 19.6\%. Homepage: \url{http://sigmod2017.org/}
\item SOCC: ACM Symposium on Cloud Computing, 2017-09-25. 45 papers, acceptance rate: unknown. Homepage: \url{https://acmsocc.github.io/2017/}
\item SOSP: Symposium on Operating Systems Principles, 2017-10-29. 39 papers, acceptance rate: 16.8\%. Homepage: \url{https://www.sigops.org/sosp/sosp17/}
\item SP: IEEE Security and Privacy, 2017-05-22. 60 papers, acceptance rate: 14.3\%. Homepage: \url{https://www.ieee-security.org/TC/SP2017/index.html}
\item SPAA: ACM Symposium on Parallelism in Algoirmths and Architectures, 2017-07-24. 31 papers, acceptance rate: 24.4\%. Homepage: \url{http://spaa.acm.org/2017/index.html}
\item SYSTOR: ACM International Systems and Storage Conference, 2017-05-22. 16 papers, acceptance rate: 34\%. Homepage: \url{https://www.systor.org/2017/}
\item VEE: ACM International Conference on Virtual Execution Environments, 2017-04-09. 18 papers, acceptance rate: 41.9\%. Homepage: \url{http://conf.researchr.org/home/vee-2017}

\end{enumerate}

\hypertarget{refs}{}
\leavevmode\hypertarget{ref-adam02:citation}{}%
Adam, David. 2002. ``Citation Analysis: The Counting House.'' \emph{Nature} 415 (6873): 726--30.

\leavevmode\hypertarget{ref-aksnes03:macro}{}%
Aksnes, Dag W. 2003. ``A Macro Study of Self-Citation.'' \emph{Scientometrics} 56 (2): 235--46. \url{https://doi.org/10.1023/A:1021919228368}.

\leavevmode\hypertarget{ref-biagioli16:watch}{}%
Biagioli, Mario. 2016. ``Watch Out for Cheats in Citation Game.'' \emph{Nature} 535 (7611): 201--1. \url{https://doi.org/10.1038/535201a}.

\leavevmode\hypertarget{ref-broch01:cite}{}%
Broch, Elana. 2001. ``Cite Me, Cite My References? (Scholarly Use of the Acm Sigir Proceedings Based on Two Citation Indexes).'' In \emph{Proceedings of the 24th Annual International Acm Sigir Conference on Research and Development in Information Retrieval}, 446--47. \url{https://doi.org/10.1145/383952.384090}.

\leavevmode\hypertarget{ref-chakraborty15:categorization}{}%
Chakraborty, Tanmoy, Suhansanu Kumar, Pawan Goyal, Niloy Ganguly, and Animesh Mukherjee. 2015. ``On the Categorization of Scientific Citation Profiles in Computer Science.'' \emph{Communications of the ACM} 58 (9): 82--90. \url{https://doi.org/10.1145/2701412}.

\leavevmode\hypertarget{ref-coupe13:peer}{}%
Coupé, Tom. 2013. ``Peer Review Versus Citations--an Analysis of Best Paper Prizes.'' \emph{Research Policy} 42 (1): 295--301. \url{https://doi.org/10.1016/j.respol.2012.05.004}.

\leavevmode\hypertarget{ref-devarakonda20:viewing}{}%
Devarakonda, Sitaram, Dmitriy Korobskiy, Tandy Warnow, and George Chacko. 2020. ``Viewing Computer Science Through Citation Analysis: Salton and Bergmark Redux.'' \emph{Scientometrics} 125 (1): 271--87. \url{https://doi.org/10.1007/s11192-020-03624-0}.

\leavevmode\hypertarget{ref-frachtenberg:github-repo}{}%
Frachtenberg, Eitan. 2021. ``Systems Conferences Analysis Dataset,'' October. \url{https://doi.org/10.5281/zenodo.5590574}.

\leavevmode\hypertarget{ref-frachtenberg22:artifact}{}%
---------. 2022a. ``Research Artifacts and Citations in Computer Systems Papers.'' \emph{PeerJ Computer Science} 8 (February): e887. \url{https://doi.org/10.7717/peerj-cs.887}.

\leavevmode\hypertarget{ref-frachtenberg22:sigmetrics}{}%
---------. 2022b. ``Multifactor Citation Analysis over Five Years: A Case Study of SIGMETRICS Papers.'' \emph{Publications} 10 (4): 47. \url{https://doi.org/10.3390/publications10040047}.

\leavevmode\hypertarget{ref-goodrum01:scholarly}{}%
Goodrum, Abby A, Katherine W McCain, Steve Lawrence, and C Lee Giles. 2001. ``Scholarly Publishing in the Internet Age: A Citation Analysis of Computer Science Literature.'' \emph{Information Processing \& Management} 37 (5): 661--75. \url{https://www.sciencedirect.com/science/article/pii/S0306457300000479}.

\leavevmode\hypertarget{ref-halevi17:suitability}{}%
Halevi, Gali, Henk Moed, and Judit Bar-Ilan. 2017. ``Suitability of Google Scholar as a Source of Scientific Information and as a Source of Data for Scientific Evaluation---Review of the Literature.'' \emph{Journal of Informetrics} 11 (3): 823--34.

\leavevmode\hypertarget{ref-hamilton91:uncited}{}%
Hamilton, David P. 1991. ``Who's Uncited Now?'' \emph{Science} 251 (4989): 25--25. \url{https://doi.org/10.1126/science.1986409}.

\leavevmode\hypertarget{ref-harter96:impact}{}%
Harter, Stephen P. 1996. ``The Impact of Electronic Journals on Scholarly Communication: A Citation Analysis.'' \emph{The Public-Access Computer Systems Review} 7 (5): 5--34.

\leavevmode\hypertarget{ref-harzing16:google}{}%
Harzing, Anne-Wil, and Satu Alakangas. 2016. ``Google Scholar, Scopus and the Web of Science: A Longitudinal and Cross-Disciplinary Comparison.'' \emph{Scientometrics} 106 (2): 787--804.

\leavevmode\hypertarget{ref-herring02:use}{}%
Herring, Susan Davis. 2002. ``Use of Electronic Resources in Scholarly Electronic Journals: A Citation Analysis.'' \emph{College \& Research Libraries} 63 (4): 334--40. \url{https://crl.acrl.org/index.php/crl/article/viewFile/15538/16984}.

\leavevmode\hypertarget{ref-heumuller20:publish}{}%
Heumüller, Robert, Sebastian Nielebock, Jacob Krüger, and Frank Ortmeier. 2020. ``Publish or Perish, but Do Not Forget Your Software Artifacts.'' \emph{Empirical Software Engineering} 25 (6): 4585--4616. \url{https://doi.org/10.1007/s10664-020-09851-6}.

\leavevmode\hypertarget{ref-hirsch05:index}{}%
Hirsch, Jorge E. 2005. ``An Index to Quantify an Individual's Scientific Research Output.'' \emph{Proceedings of the National Academy of Sciences} 102 (46): 16569--72. \url{https://doi.org/10.1073/pnas.0507655102}.

\leavevmode\hypertarget{ref-hirst77:computer}{}%
Hirst, Graeme, and Nadia Talent. 1977. ``Computer Science Journals---an Iterated Citation Analysis.'' \emph{IEEE Transactions on Professional Communication} PC-20 (4): 233--38. \url{https://doi.org/10.1109/TPC.1977.6591956}.

\leavevmode\hypertarget{ref-iqbal19:sigcomm}{}%
Iqbal, Waleed, Junaid Qadir, Saeed-Ul Hassan, Rana Tallal Javed, Adnan Noor Mian, Jon Crowcroft, and Gareth Tyson. 2019. ``Five Decades of the Acm Special Interest Group on Data Communications (Sigcomm) a Bibliometric Perspective.'' \emph{ACM SIGCOMM Computer Communication Review} 49 (5): 29--37. \url{https://doi.org/10.1145/3371934.3371948}.

\leavevmode\hypertarget{ref-iqbal19:bibliometric}{}%
Iqbal, Waleed, Junaid Qadir, Gareth Tyson, Adnan Noor Mian, Saeed-ul Hassan, and Jon Crowcroft. 2019. ``A Bibliometric Analysis of Publications in Computer Networking Research.'' \emph{Scientometrics} 119 (2): 1121--55. \url{https://doi.org/10.1007/s11192-019-03086-z}.

\leavevmode\hypertarget{ref-jacques10:impact}{}%
Jacques, Thomas S, and Neil J Sebire. 2010. ``The Impact of Article Titles on Citation Hits: An Analysis of General and Specialist Medical Journals.'' \emph{JRSM Short Reports} 1 (1): 1--5. \url{https://doi.org/10.1258/shorts.2009.100020}.

\leavevmode\hypertarget{ref-khan22:cloud}{}%
Khan, Daud, Masoumeh Khalil Arjmandi, and Mayank Yuvaraj. 2022. ``Most Cited Works on Cloud Computing: The `Citation Classics' as Viewed Through Dimensions.ai.'' \emph{Science \& Technology Libraries} 41 (1): 42--55. \url{https://doi.org/10.1080/0194262X.2021.1951424}.

\leavevmode\hypertarget{ref-kousha07:google}{}%
Kousha, Kayvan, and Mike Thelwall. 2007. ``Google Scholar Citations and Google Web/Url Citations: A Multi-Discipline Exploratory Analysis.'' \emph{Journal of the American Society for Information Science and Technology} 58 (7): 1055--65.

\leavevmode\hypertarget{ref-lariviere09:decline}{}%
Larivière, Vincent, Yves Gingras, and Éric Archambault. 2009. ``The Decline in the Concentration of Citations, 1900--2007.'' \emph{Journal of the American Society for Information Science and Technology} 60 (4): 858--62. \url{https://arxiv.org/ftp/arxiv/papers/0809/0809.5250.pdf}.

\leavevmode\hypertarget{ref-lariviere16:simple}{}%
Larivière, Vincent, Veronique Kiermer, Catriona J MacCallum, Marcia McNutt, Mark Patterson, Bernd Pulverer, Sowmya Swaminathan, Stuart Taylor, and Stephen Curry. 2016. ``A Simple Proposal for the Publication of Journal Citation Distributions.'' \emph{BioRxiv}, 062109.

\leavevmode\hypertarget{ref-lee19:predictive}{}%
Lee, Danielle H. 2019. ``Predictive Power of Conference-Related Factors on Citation Rates of Conference Papers.'' \emph{Scientometrics} 118 (1): 281--304. \url{https://doi.org/10.1007/s11192-018-2943-z}.

\leavevmode\hypertarget{ref-lister08:citation}{}%
Lister, Raymond, and Ilona Box. 2008. ``A Citation Analysis of the Sigcse 2007 Proceedings.'' In \emph{Proceedings of the 39th Sigcse Technical Symposium on Computer Science Education}, 476--80.

\leavevmode\hypertarget{ref-martin18:google}{}%
Martin-Martin, Alberto, Enrique Orduna-Malea, Mike Thelwall, and Emilio Delgado Lopez-Cozar. 2018. ``Google Scholar, Web of Science, and Scopus: A Systematic Comparison of Citations in 252 Subject Categories.'' \emph{Journal of Informetrics} 12 (4): 1160--77.

\leavevmode\hypertarget{ref-martins10:assessing}{}%
Martins, Waister, Marcos Gonçalves, Alberto Laender, and Nivio Ziviani. 2010. ``Assessing the Quality of Scientific Conferences Based on Bibliographic Citations.'' \emph{Scientometrics} 83 (1): 133--55.

\leavevmode\hypertarget{ref-mattauch20:bibliometric}{}%
Mattauch, Sandra, Katja Lohmann, Frank Hannig, Daniel Lohmann, and Jürgen Teich. 2020. ``A Bibliometric Approach for Detecting the Gender Gap in Computer Science.'' \emph{Communications of the ACM} 63: 74--80. \url{https://doi.org/10.1145/3376901}.

\leavevmode\hypertarget{ref-meho07:rise}{}%
Meho, Lokman I. 2007. ``The Rise and Rise of Citation Analysis.'' \emph{Physics World} 20 (1): 32. \url{https://arxiv.org/pdf/physics/0701012,}

\leavevmode\hypertarget{ref-moed06:citation}{}%
Moed, Henk F. 2006. \emph{Citation Analysis in Research Evaluation}. Springer Science \& Business Media.

\leavevmode\hypertarget{ref-parthasarathy14:sentiment}{}%
Parthasarathy, G, and DC Tomar. 2014. ``Sentiment Analyzer: Analysis of Journal Citations from Citation Databases.'' In \emph{5th International Conference-Confluence the Next Generation Information Technology Summit (Confluence)}, 923--28. IEEE. \url{https://doi.org/10.1109/CONFLUENCE.2014.6949321}.

\leavevmode\hypertarget{ref-patience17:citation}{}%
Patience, Gregory S., Christian A. Patience, Bruno Blais, and Francois Bertrand. 2017. ``Citation Analysis of Scientific Categories.'' \emph{Heliyon} 3 (5): e00300. \url{https://doi.org/https://doi.org/10.1016/j.heliyon.2017.e00300}.

\leavevmode\hypertarget{ref-patterson04:health}{}%
Patterson, David A. 2004. ``The Health of Research Conferences and the Dearth of Big Idea Papers.'' \emph{Communications of the ACM} 47 (12): 23--24. \url{https://doi.org/10.1145/1035134.1035153}.

\leavevmode\hypertarget{ref-patterson99:evaluating}{}%
Patterson, David A, Lawrence Snyder, and Jeffrey Ullman. 1999. ``Evaluating Computer Scientists and Engineers for Promotion and Tenure.'' \emph{Computing Research News}, September. \url{http://www.cra.org/resources/bp-view/evaluating_computer_scientists_and_engineers_for_promotion_and_tenure/}.

\leavevmode\hypertarget{ref-pichappan99:skewness}{}%
Pichappan, P, and R Ponnudurai. 1999. ``Skewness in Citation Peak.'' In \emph{Proceedings of the 7th Conference of the International Society for Scientometrics and Informetrics}, 395--407.

\leavevmode\hypertarget{ref-rahm05:citation}{}%
Rahm, Erhard, and Andreas Thor. 2005. ``Citation Analysis of Database Publications.'' \emph{ACM Sigmod Record} 34 (4): 48--53. \url{https://doi.org/10.1145/1107499.1107505}.

\leavevmode\hypertarget{ref-redner98:popular}{}%
Redner, Sidney. 1998. ``How Popular Is Your Paper? An Empirical Study of the Citation Distribution.'' \emph{The European Physical Journal B-Condensed Matter and Complex Systems} 4 (2): 131--34. \url{https://link.springer.com/content/pdf/10.1007/s100510050359.pdf}.

\leavevmode\hypertarget{ref-snyder98:patterns}{}%
Snyder, Herbert, and Susan Bonzi. 1998. ``Patterns of Self-Citation Across Disciplines (1980-1989).'' \emph{Journal of Information Science} 24 (6): 431--35. \url{https://doi.org/10.1177/016555159802400606}.

\leavevmode\hypertarget{ref-tsay11:journal}{}%
Tsay, Ming-yueh, and Zhu-yee Shu. 2011. ``Journal Bibliometric Analysis: A Case Study on the Journal of Documentation.'' \emph{Journal of Documentation} 67 (5): 806--22. \url{https://doi.org/10.1108/00220411111164682}.

\leavevmode\hypertarget{ref-varga19:shorter}{}%
Varga, Attila. 2019. ``Shorter Distances Between Papers over Time Are Due to More Cross-Field References and Increased Citation Rate to Higher-Impact Papers.'' \emph{Proceedings of the National Academy of Sciences} 116 (44): 22094--9. \url{https://doi.org/10.1073/pnas.1905819116}.

\leavevmode\hypertarget{ref-vrettas15:conferences}{}%
Vrettas, George, and Mark Sanderson. 2015. ``Conferences Versus Journals in Computer Science.'' \emph{Journal of the Association for Information Science and Technology} 66 (12): 2674--84. \url{https://doi.org/10.1002/asi.23349}.

\leavevmode\hypertarget{ref-wainer15:peer}{}%
Wainer, Jacques, Michael Eckmann, and Anderson Rocha. 2015. ``Peer-Selected `Best Papers'---Are They Really That `Good'?'' \emph{PLOS ONE} 10 (3): e0118446. \url{https://doi.org/10.1371/journal.pone.0118446}.

\leavevmode\hypertarget{ref-waltman16:review}{}%
Waltman, Ludo. 2016. ``A Review of the Literature on Citation Impact Indicators.'' \emph{Journal of Informetrics} 10 (2): 365--91. \url{https://doi.org//10.1016/j.joi.2016.02.007}.

\leavevmode\hypertarget{ref-wang21:science}{}%
Wang, Dashun, and Albert-László Barabási. 2021. \emph{The Science of Science}. Cambridge, UK: Cambridge University Press.

\leavevmode\hypertarget{ref-wang13:citation}{}%
Wang, Jian. 2013. ``Citation Time Window Choice for Research Impact Evaluation.'' \emph{Scientometrics} 94 (3): 851--72.

\leavevmode\hypertarget{ref-wang16:cloud}{}%
Wang, Nianxin, Huigang Liang, Yu Jia, Shilun Ge, Yajiong Xue, and Zhining Wang. 2016. ``Cloud Computing Research in the Is Discipline: A Citation/Co-Citation Analysis.'' \emph{Decision Support Systems} 86: 35--47.

\leavevmode\hypertarget{ref-wolfgang04:bibliometric}{}%
Wolfgang, Glänzel, Thijs Bart, and Schlemmer Balázs. 2004. ``A Bibliometric Approach to the Role of Author Self-Citations in Scientific Communication.'' \emph{Scientometrics} 59 (1): 63--77. \url{https://doi.org/10.1023/b:scie.0000013299.38210.74}.

\leavevmode\hypertarget{ref-wu09:research}{}%
Wu, Ling-Ling, Luesak Luesukprasert, and Lynne Lee. 2009. ``Research and the Long Tail: A Large-Scale Citation Analysis.'' In \emph{42nd Hawaii International Conference on System Sciences}, 1--10. IEEE. \url{https://ieeexplore.ieee.org/stamp/stamp.jsp?arnumber=4755756}.

\bibliographystyle{unsrt}

\end{document}